\documentclass[aps,superscriptaddress,showpacs,nofootinbib,eqsecnum]{revtex4}

\usepackage{eucal}
\usepackage{epsfig}

\begin{document}

\title{An Effective Field Theory Model for One-Dimensional ${\rm CH}$ Chains: Effects at Finite Chemical Potential, Temperature\\
and External Zeeman Magnetic Field}

\author{Heron Caldas} \email{hcaldas@ufsj.edu.br} \affiliation{Departamento de
  Ci\^{e}ncias Naturais. Universidade Federal de S\~{a}o Jo\~{a}o del Rei,\\
  36301-160, S\~{a}o Jo\~{a}o del Rei, MG, Brazil}

\begin{abstract}

In this work we use an effective field theory model to investigate doped $\rm (CH)_x$ chains under the influence of an external constant Zeeman magnetic field $B_0$, at zero and finite temperature, in the mean-field approximation and beyond. We consider both homogeneous and inhomogeneous $\Delta(x)$ condensates and calculate the Pauli magnetization and the magnetic susceptibility of these chains at various situations for the temperature and chemical potentials. We also briefly discuss the possibility of using these materials as partially polarized 1D organic conductors.

\end{abstract}


\maketitle


\section{Introduction}

In the last few years graphene has attracted the attention of both theoretical and experimental communities due to the interesting features this two-dimensional system presents. The electrons of this material are Dirac-like, with linear dispersion relations near the $K$ points. Monolayer graphene can be cut in stripes to construct one-dimensional (1D) graphene wires, which have been studied in various recent works~\cite{Novikov,Silvestrov}. In Ref.~\cite{Faccioli}, an effective field theory has been used to investigate the low-energy quantum electrodynamics of Dirac electrons in an undoped graphene wire.

Another organic material which presents similarities with a graphene wire is a $ \rm CH$ chain, which for more than 3 decades have also created an admiration due to the unexpected and fascinating discover that doped {\it trans}-polyacetylene (TPA), a 1D polymer, behaves as a metal, exhibiting electrical conductivity of some metals, like copper~\cite{paper1}. Since then, several promising properties, such as electronic, optical and magnetic, have been shown by conjugated polymers, which give these materials the possibility of application in the semiconductor nanotechnology~\cite{Nature1,Aleshin}.

Polyacetylene is a linear chain of $\rm CH$ groups which can have two forms, {\it trans} and {\it cis}. The {\it trans} form, is more stable and, in the absence of doping, has a doubly-degenerate ground state. The {\it trans}-${\rm (CH)_x}$ has one $\pi-$electron per site, and would be a metal in the tight binding approximation. However, spontaneous Peierls dimerization (i.e., the interaction of the electrons with the lattice) turns TPA into an insulator. After the critical doping, the conductivity of TPA is greatly increased, reaching metal-like properties~\cite{Review}. Experiments show that the observed non-metal to metal first-order transition in polyacetylene happens when the dopant concentration $y$, defined as the number of doped electrons per $100$ carbon atoms, is increased up to a critical value $y_c \cong 6$~\cite{Fernando}. Recent calculations show that thermal effects have a weak effect on $y_c$~\cite{PRB}.

The electron-phonon interactions in $\rm CH$ is successful described by the Su-Schrieffer-Heeger (SSH) Hamiltonian~\cite{SSH}, whose continuum version is known as the Takayama--Lin-Liu--Maki (TLM) model \cite{TLM}. The TLM model is a relativistic field theory with two-flavor Dirac fermions. We shall use the Gross-Neveu (GN) model \cite{GN} in 1+1 dimensions since its Lagrangian is equivalent to that of the TLM model \cite{CB1,CB2}. The GN model is a well-known 1D theory with four-fermion interactions. The choice of employing an effective quantum field theory model (QFTM) to the investigation we do in the present work is basically motivated by the analytical and simple solutions the GN model provides, as we will see below. Besides, one of the most important features of the GN model which enables its use as an effective theory to describe materials that undergo transitions from non-metallic to metallic phases, as the polymers we are considering, is that it is asymptotically free, just like QCD. This means that the fermions become non-interacting at high enough densities and are not able to support their condensate that they formed to break chiral symmetry at zero or very low densities. Therefore, since the theory displays a phase transition at a certain fermion density or at the corresponding value of chemical potential, the use of the GN model is very appropriate in this context. In other words, or in the field theory language, the discrete chiral symmetry is spontaneously broken in the GN model and there is a dynamical generation of a mass (gap) for the electrons. A global chemical potential $\mu$ is introduced in the theory to represent the extra electrons that are inserted in the system by the doping process. At a critical chemical potential $\mu_c=\frac{\Delta_0}{\sqrt{2}}$~\cite{Wolff}, where $\Delta_0$ ($\approx 0.7 {\rm eV}$ for {\it trans}-$(\rm CH)_x$) is the constant band-gap, the GN model undergoes a first order phase transition to a symmetry restored (zero gap) phase, agreeing with the experimentally observed non-metal to metal first-order transition in TPA~\cite{Fernando}. Although some times we shall refer to TPA throughout the paper (due to the large amount of data it has), the main results are applicable to any 1D chain with the same structure of $(\rm CH)_x$.

The critical doping concentration can be related with the critical chemical potential as $y_c=\frac{N}{\pi \hbar v_F} a \mu_c$~\cite{CM}, where $N=2$ can be regarded as an internal number (spin) of fermion flavors, $\hbar$ is the Planck's constant divided by $2\pi$, $v_F=k_F\hbar/m$ is the Fermi velocity, $k_F$ is the Fermi wavenumber, $m \equiv \hbar^2/2t_o a^2$, and $a$ ($\cong 1.22 {\rm \AA} $ for {\it trans}-$(\rm CH)_x)$) is the lattice (equilibrium) spacing between the $x$ coordinates of successive ${\rm CH}$ radicals in the undimerized structure, and $t_0 (\approx 3~ {\rm eV}$ for {\it trans}-$(\rm CH)_x)$~\cite{Review} is the intercarbon transfer integral for $\pi$ electrons, known as the ``hopping parameter". In Ref.~\cite{CM} the GN model has been used in the large-$N$ (mean-field) approximation as an effective model to describe the non-metal to metal phase transition in TPA, and they found a very good agreement with the experimentally measured $y_c (\cong 6 \%)$.

In this work we use the GN model as a QFTM for doped $\rm (CH)_x$ chains to investigate the consequences of a $B_0$ applied parallel to these ``wires" at zero and finite temperature ($T$). We show that the actuation of $B_0$ produces an imbalance between the chemical potentials (and consequently in the densities) of the spin-up ($\uparrow$) and spin-down ($\downarrow$) conduction electrons inserted in the system. Depending on the level of imbalance between the chemical potentials of the two involved fermionic species the $\rm (CH)_x$ chain can be in a non-condensed (zero gap) partially polarized state, or can be turned into a fully polarized state with a small but non-zero gap~\cite{Caldas2}. It is shown that in {\it trans}-$\rm (CH)_x$ with low imbalance (with respect to $\mu_c$), at zero $T$, the system acquires a partial spin polarization. With strong imbalance there is total absence of electrons from one of the two possible spin orientations ($\uparrow$ or $\downarrow$), meaning that the system is now fully polarized~\cite{Caldas2}. We expect that this might have observable consequences in the magnetic properties of doped $\rm (CH)_x$ or other conducting polymers. In the investigation of spin-polarized phases and the magnetic properties of 1D {\it cis}-polymers, one has to consider the massive Gross-Neveu model~\cite{Aragao}, as we shall do here.

It is important to notice that the study of the effects of an external constant Zeeman magnetic field $B_0$ applied on a physical system is relevant in many physical situations, as in the investigation of metal-insulator transition~\cite{Dent} and magnetization~\cite{PRB2} in 2D electron systems.

The paper is organized as follows. In Section II we introduce the discrete and continuum model Hamiltonians and the model Lagrangian describing $\rm (CH)_x$ chains. In Section III the temperature dependent renormalized effective potential is obtained in the mean-field approximation and beyond. In this section we obtain an analytical expression for the effective potential at high temperature, as well as the chemical potentials dependent gap equation and the critical temperature at which the gap vanishes. Besides, the tricritical points of the second order transition line of the gap parameter are explicitly calculated by considering both homogeneous and inhomogeneous gap parameters. The temperature dependent magnetic properties of $\rm (CH)_x$ chains are also obtained in this section. For completeness we study in Section IV the Massive Gross-Neveu model which is appropriate to describe Cis-Polyacetylene. In Section V we show a Summary of the main results and finally, we present the conclusions in Section VI.
 
\section{The Discrete and Continuum Models}

\subsection{The Model Hamiltonians}

\subsubsection{The SSH Discrete Model Hamiltonian}

The tight-binding SSH Hamiltonian with electron-phonon interactions has the following form~\cite{SSH}:

\begin{equation}
\label{H1}
H = H_{\pi}  + H_{ph}+ H_{\pi-ph},
\end{equation}
where 

\begin{equation}
H_{\pi}= -t_0 \sum_{n,s} (c_{n+1,s}^\dagger c_{n,s} +  c_{n,s}^\dagger c_{n+1,s}),
\label{H2}
\end{equation}
describes the $\pi-$electron (${\it e}$) hopping between site $n$ and $n+1$. $t_0$ is the hopping constant for an undimerized structure,

\begin{equation}
H_{ph}=\sum_{n} \frac {p_n^2} {2M_{\rm CH}}+ \frac {K} {2} \sum_{n}(u_{n+1}- u_n)^2,
\label{H3}
\end{equation}
represents the lattice, i.e. the phonons (${\it ph}$), $M_{\rm CH}$ is the mass of the ${\rm CH}$ group, $K$ is the ``spring'' constant, $u_n$ is the deviation from the undimerized structure on site $n$, and the interactions between these two is

\begin{equation}
H_{\pi-ph}=\alpha \sum_{n,s} (u_{n+1}- u_n)(c_{n+1,s}^\dagger c_{n,s} +  c_{n,s}^\dagger c_{n+1,s}),
\label{H4}
\end{equation}
where $\alpha$ is the ${\it e-ph}$ interaction constant. Summing $H_{\pi}$ and $H_{\pi-ph}$ one defines $t_{n,n+1}=t_0 - \alpha (u_{n+1}-u_n)$.

\subsubsection{Electron Correlations}

Electron correlations i.e., electron-electron interactions, are fundamental in the explanation of many physical phenomena displayed by conducting polymers as, for example, the observed negative spin densities in ${\it trans}$-$\rm (CH)_x$, which demonstrates electronic correlations in these chains~\cite{EC1}. Since the SSH model (that is based only on electron-phonon interactions) gives zero spin density, other models that take into account electron correlations are naturally necessary. Besides, the inclusion of electron correlations are mandatory for the optical spectra of even Polyenes~\cite{Soos1}. 

The addition of a term describing electron interactions in the SSH Hamiltonian defines the Pariser-Parr-Pople (PPP) models~\cite{Soos1,Soos2}, which provides an extension of the Hubbard model by properly accounting for the long-range character of the electron-electron (Coulomb) repulsion. Thus, we have:

\begin{equation}
\label{PPP1}
H_{PPP} = H + H_{\pi-\pi},
\end{equation}
where $H$ is the noninteracting Hamiltonian defined in Eq.~(\ref{H1}) and $H_{\pi-\pi}$ is given by

\begin{equation}
\label{PPP2}
H_{\pi-\pi}=\frac{1}{2} U \sum_{l} n_{l} n_{l} + \frac{1}{2} \sum_{l, l'} V_{l,l'} n_{l} n_{l'},
\end{equation}
where $U$ is an on-site Hubbard term, $V_{l,l'}$ is an off-site interactions between electrons on site $l$ and $l'$, and $n_{l} =c_{l,\alpha}^\dagger  c_{l, \alpha} + c_{l,\beta}^\dagger  c_{l, \beta} - 1$ is the net charge density on site $l$ with spin $\alpha$ ($\beta$). The prime in the last term implies $l \neq l'$. The on-site correlations $U$, and inter-site
Coulomb interactions $V_{l,l'}$ are, in principle, arbitrary, and generally the Ohno's formula~\cite{Ohno} is employed~\cite{Soos2,Wu}:

\begin{equation}
\label{PPP3}
V_{l,l'}= \frac{U}{\sqrt{1+\left(\frac{r_{l,l'}}{r_0} \right)^2}},
\end{equation}
where $r_0 \approx  1.29 \AA$, $r_{l,l'}$ is the distance between sites $l$ and $l'$ in units of ${\rm \AA}$, and $U = 11.13-11.26~ {\rm eV}$. Thus, as pointed out in Ref.~\cite{Soos2}, the geometry and $U$ fix all inter-site interactions.

The influence of electronic correlations on structural and electronic properties of ${\it trans}$-$\rm (CH)_x$, such as the ground-state energy, the amplitude of bond alternation and the effective force constant for the bond-stretching mode, has been investigated using the Hamiltonian of Eq.~(\ref{PPP1}) with $V_{l,l'}=0$ but considering an effective $U$~\cite{Maki}. The study of Ref.~\cite{Maki} finds a consistent modeling of ${\it trans}$-$\rm (CH)_x$ if $U$ is chosen to be of the order of $7-9~{\rm eV}$.

Since the consideration of electron correlations is out of the scope of the present work, in what follows we consider $U=V_{l,l'}=0$.

\subsubsection{The TLM Continuum Model Hamiltonian}

As pointed out in Ref.~\cite{TLM}, in the weak coupling limit, only electrons of the Fermi surface with energy of order $\approx \Delta$ (the gap energy) are affected by the dimerization. Then it is reasonable to linearize the band structure in the vicinity of the points $\pm k_F$, where $k_F$ is the Fermi momentum. After the linearization, TLM got the continuum version of (\ref{H1}):

\begin{eqnarray}
\label{eq-214}
H = \frac {1} {2\pi\hbar v_F \lambda_{TLM}} \int dx\left[\Delta^2(x) + \frac{\dot{\Delta}^2}{\Omega^2_0}\right ]
+ \sum_s \int dx~{\psi^s}^{\dagger}(x) [-i\hbar v_F\sigma_3\partial_x + \Delta(x)\sigma_1] \psi^s (x),
\end{eqnarray}
where the constants above are related with those of the original SSH Hamiltonian as $\hbar v_F=2t_0a, \lambda_{TLM} =(2\alpha)^2/2\pi Kt_0$, $\Omega^2_0=4K/M_{\rm CH}$ is a dimensionless coupling, $\sigma_x$ are the Pauli matrices, $\Delta(x)$ is a real gap related to lattice vibrations, and $\dot \Delta(x) \equiv d \Delta(x)/dt$. $\psi^s$ is a two field component spinor $\psi^s= \left(\begin{array}{cc} \psi_{L}^s\\ \psi_{R}^s \end{array} \right)$, representing the ``left moving" and ``right moving" electrons close to their Fermi energy, respectively, where $s$ is an internal symmetry index (spin) that determines the effective degeneracy of the fermions, $s=1 \equiv \uparrow$, and $s= 2\equiv \downarrow$.

\subsection{The Model Lagrangian}

The Lagrangian density of the TLM model in the adiabatic approximation ($\dot \Delta(x)=0$), is given by

\begin{eqnarray}
{\cal L}_{\rm TLM} =
\sum_s {\psi^s}^{\dagger}
\left( i \hbar \partial_t - i \hbar v_F \gamma_5 \partial_x - \gamma_0
\Delta(x) 
\right) \psi^s
-\frac{1}{2 \pi \hbar v_F \lambda_{\rm TLM}} 
\Delta^2(x)\;,
\label{LagTLM}
\end{eqnarray}
where the gamma matrices are given in terms of the Pauli matrices, as $\gamma_0=\sigma_1$, and $\gamma_5=-\sigma_3$.

The four-fermion Lagrangian density of the massless GN model~\cite{GN} reads

\begin{eqnarray}
{\cal L}_{\rm GN} =
\sum_s \bar \psi^{s} \left( i \hbar \gamma_0 \partial_t - i \hbar v_F \gamma_1 \partial_x \right) \psi^s  
+ \frac{ \lambda_{\rm GN } \hbar v_F}{2 N} \left( \bar\psi \psi  \right)^2 \;,
\label{LagGN}
\end{eqnarray}
where $\bar \psi =  {\psi}^{\dagger} \gamma_0$ and $\bar\psi \psi = \bar\psi^s \psi^s = \sum_{s=1}^{N} \bar\psi^s \psi^s$, and $N$ is the number of flavors of Dirac fermions in the effective field theory. In the large-{N} (mean-field) approximation, one obtains~\cite{GN,PRB}

\begin{eqnarray}
{\cal L}_{\rm GN} =
\sum_s {\psi^s}^{\dagger}
\left( i  \hbar \partial_t - i \hbar v_F \gamma_5 \partial_x - \gamma_0
\Delta(x) 
\right) \psi^s 
-\frac{N}{2  \hbar v_F \lambda_{\rm GN}} 
\Delta^2(x)\;.
\label{LagGN2}
\end{eqnarray}
Thus one sees that the equivalence between the TLM and the Gross-Neveu (GN) model is established by setting $\lambda_{\rm TLM} = \frac{\lambda_{\rm GN}}{N \pi} $. Note that the fermion-lattice interaction in the equation above resembles the usual fermion-boson interaction in quantum field theory context~\cite{NPB1}. 

In order to consider the application of an external Zeeman magnetic field $B_0$ parallel to the system and its effects, it is convenient to start by writing the grand canonical partition function associated with ${\cal L}_{\rm GN}$ in the imaginary time formalism~\cite{Kapusta}:

\begin{equation}
\label{action1}
{\cal Z}= \int  ~ {\cal D} \bar \psi ~ {\cal D} \psi~ exp \left\{ \int_0^{ \beta} d\tau \int dx~ \left[ { L}_{\rm GN} \right] \right\},
\end{equation}
where $\beta \equiv 1/k_BT$, $k_B$ is the Boltzmann constant, and $L_{\rm GN}$ is the Euclidean GN Lagrangian density:

\begin{equation}
\label{L1}
{ L}_{\rm GN}= \sum_{s=1,2} \bar \psi^s [- \gamma_0 \hbar \partial_\tau + i \hbar v_F \gamma_1 \partial_x -  \Delta(x) +  \gamma_0 \mu + \gamma_0 S_s g \mu_B B_0  ] \psi^s -\frac{1}{ \hbar v_F \lambda_{\rm GN}}  \Delta^2(x),
\end{equation}
where $\mu$ is the chemical potential, $S_s=\pm 1/2$, $g$ is the effective $g$-factor and $\mu_B=e \hbar/2m \approx  5.788 \times 10^{-5} ~{\rm eV}~ {\rm T}^{-1}$ is the Bohr magneton. The parallel magnetic field couples to the electrons' spin and produces the Zeeman splitting energy term $\Delta E= S_s g \mu_B B_0$~\cite{Madelung,Kittel} in Eq.~(\ref{L1}). {}From the form of the Zeeman energy term in Eq. (\ref{L1}) we see that it can be added to the chemical potential, thus defining an effective chemical potential terms in the Lagrangian density of the form,

\begin{eqnarray}
\sum_{s=1,2} \mu_s \bar{\psi}^s \gamma_0 \psi^s &=&
\sum_{s=1,2}  \left(\mu + S_s \, g \, \mu_B
B_0\right) \bar{\psi}^s \gamma_0 \psi^s \nonumber \\ &=&
\mu_{\uparrow} {\psi_\uparrow}^{\dagger}  \psi_{\uparrow} + \mu_{\downarrow}
   {\psi_\downarrow}^{\dagger}  \psi_{\downarrow}\;,
\label{asym}
\end{eqnarray}
where $\mu_\uparrow= \mu + \delta \mu$, $\mu_\downarrow=  \mu - \delta \mu$, and $\delta \mu \equiv \frac{g}{2} \mu_B B_0$. In \cite{NPB1} it has been chosen $ \mu=\mu_c$, such that when $\delta \mu = B_0=0$ the system is in the symmetry restored phase. Thus we rewrite Eq.~(\ref{L1}) as

\begin{equation}
\label{L1Rew}
{ L}_{\rm GN}= \sum_{s=1,2} \bar \psi^s [- \gamma_0 \hbar \partial_\tau + i \hbar v_F \gamma_1 \partial_x -  \Delta(x) +  \gamma_0 \mu_s ] \psi^s -\frac{1}{ \hbar v_F \lambda_{\rm GN}}  \Delta^2(x),
\end{equation}

Integrating over the fermion fields leads to 

\begin{equation}
\label{action2}
{\cal Z} =  exp~{ \left\{ -\frac{\beta}{ \hbar v_F \lambda_{\rm GN}} \int dx ~ \Delta^2(x) \right\}} ~\Pi_{j=1}^2~det D_j ,
\end{equation}
where $D_j=- \gamma_0 \partial_\tau + i \hbar v_F \gamma_1 \partial_x +  \gamma_0 \mu_j -  \Delta(x) $ is the Dirac operator at finite temperature and density. Since $\Delta(x)$ is static, one can transform $D_j$ to the $\omega_n$ plane, where $\omega_n=(2n+1)\pi T$ are the Matsubara
frequencies for fermions, yielding $D_j= (- i \omega_n + \mu_j)\gamma_0 + i \hbar v_F \gamma_1 \partial_x  -  \Delta(x) $. After using an elementary identity $\ln (det(D_j))={\rm Tr} \ln(D_j)$, one can define the bare effective action for the static $\Delta(x)$ condensate

\begin{equation}
\label{action3}
S_{eff} [\Delta]= -\frac{\beta}{ \hbar v_F \lambda_{\rm GN}}  \int dx ~ \Delta^2(x) + \sum_{j=1}^2 {{\rm Tr} \ln (D_j)},
\end{equation}
where the trace is to be taken over both Dirac and functional indices. The condition to find the stationary points of $S_{eff} [\Delta]$ reads

\begin{equation}
\label{action4}
\frac{\delta S_{eff} [\Delta]}{\delta \Delta(x)}=0= -\frac{2 \beta}{ \hbar v_F \lambda_{\rm GN}} \Delta(x) + \frac{\delta}{\delta \Delta(x)} \left[ \sum_{j=1}^2 {{\rm Tr} \ln (D_j)} \right].
\end{equation}
The equation above is a complicated and generally unknown functional equation for $\Delta(x)$, whose solution has been investigated at various times in the literature~\cite{Dashen,Feinberg,gnpolymers,gnpolymers2,gnpolymers3,Basar}. Its solution is not only of academic interest, but has direct application in condensed matter physics as, for example, in~\cite{Review,gnpolymers3}, and in the present work.

\section{Symmetry break and Condensates at Mean-Field and Beyond}
\label{SB}

The GN model in 1D has a discrete chiral symmetry which is dynamically broken by the non-perturbative vacuum~\cite{Dashen}. As a result, there will be the generation of homogeneous and inhomogeneous $\Delta(x)$ condensates, as discussed below.

\subsection{Homogeneous $\Delta(x)$ Condensates}
\label{homogeneous}

For a constant $\Delta$ field the Dirac operator reads $D_j= (- i \omega_n + \mu_j)\gamma_0 + i \hbar v_F \gamma_1 p  -  \Delta $, so the trace in Eq.~(\ref{action3}) can be evaluated in a closed form for the asymmetrical ($\delta \mu \neq 0$) system~\cite{NPB1}. From Eq.~(\ref{action2}) one obtains the ``effective" potential $V_{eff}=-\frac{k_B T}{L} \ln {\cal Z}$, where $L$ is the length of the system:

\begin{eqnarray}
\label{poteff}
V_{eff}(\Delta,\mu_{\uparrow,\downarrow},T)=\frac{1}{ \hbar v_F \lambda_{\rm GN}} \Delta^2 
&-& k_B T \int^{+\infty}_{-\infty}{\frac{dp}{2\pi \hbar}}~ \Big[ 2 \beta E_p
+  \ln \left(1+e^{-\beta E_\uparrow^+}\right)+  \ln \left(1+e^{-\beta E_\uparrow^-}\right)\\
\nonumber
&+& \ln \left(1+e^{-\beta E_\downarrow^+}\right) + \ln \left(1+e^{-\beta E_\downarrow^-}\right) \Big],
\end{eqnarray}
where $E_{\uparrow,\downarrow}^{\pm} \equiv E_p \pm \mu_{\uparrow,\downarrow}$, $E_p = \sqrt{v_F^2 p^2+\Delta^2}$.

\subsubsection{Zero Temperature and Zero Chemical Potentials}

The first term in the integration in $p$ in Eq.~(\ref{poteff}), corresponding to the vacuum part ($\mu_{\uparrow,\downarrow}=T=0$), is divergent. Introducing a momentum cutoff $\Lambda$ to regulate this part of $V_{eff}$, one obtains, after renormalization where the momentum cutoff is taken to infinity while maintaining finite and stable results, the following expression to the effective potential~\cite{GN,Caldas2}:

\begin{eqnarray}
V_{eff}(\Delta)=\frac{\Delta^2}{ \hbar v_F } \left(\frac{1}{\lambda_{\rm GN}} - \frac{3}{2 \pi} \right) + \frac{\Delta^2}{\pi \hbar v_F } \ln \left( \frac{\Delta}{m_F} \right),
\label{poteff2}
\end{eqnarray}
where $m_F$ is an arbitrary renormalization scale, with dimension of energy, introduced during the
regularization process used to compute the appropriate momentum integrals. The minimization of $V_{eff}(\Delta)$ with respect to $\Delta$ gives $\Delta_0=0$ and the well-known result for the non-trivial gap~\cite{GN}:

\begin{equation}
\Delta_0 = m_F e^{ 1- \frac {\pi}{\lambda_{\rm GN} }}.
\label{delta0}
\end{equation}
This is the phenomenon of dynamical mass generation in the massless model. From this equation it is easy to see that with the experimentally measured $\Delta_0$ and $\alpha$, $t_0$ and $K$ which enters $\lambda_{\rm TLM}$, one sets the value of $m_F$. Equation~(\ref{poteff2}) can be expressed in a more convenient form in terms of $\Delta_0$ as

\begin{eqnarray}
\label{poteff2_1}
V_{eff}(\Delta)=\frac{\Delta^2}{2 \pi \hbar v_F } \left[ \ln \left( \frac{\Delta^2}{\Delta_0^2} \right)-1 \right],
\end{eqnarray}
which is clearly symmetric under $\Delta \to -\Delta$, that generates the discrete chiral symmetry of the GN model. As we pointed out before this discrete symmetry is dynamically broken and thus there is a kink solution interpolating between the two degenerate minima $\Delta = \pm \Delta_0$ of (\ref{poteff2_1}) at $x = \pm \infty$~\cite{Zee}:

\begin{equation}
\Delta(x)=\Delta_0 \tanh(\Delta_0 x).
\label{kink}
\end{equation}
In the next subsection we discuss the effects of space dependent $\Delta(x)$ condensates. 

\subsubsection{Zero Temperature and Finite Chemical Potentials}

At finite chemical potentials and in the zero temperature limit, Eq.~(\ref{poteff}) reads

\begin{eqnarray}
V_{eff}(\Delta,\mu_{\uparrow,\downarrow})=\frac{1}{ \hbar v_F \lambda_{\rm GN}} \Delta^2 -  2 \int^{\Lambda}_0{\frac{dp}{\pi \hbar}}~  E_p
+\int_0^{p_{F}^\uparrow} \frac{dp}{\pi \hbar} E_\uparrow^-  + \int_0^{p_{F}^\downarrow} \frac{dp}{\pi \hbar} E_\downarrow^-,
\label{poteff-4}
\end{eqnarray}
where $p_{F}^{\uparrow,\downarrow}=\frac{1}{v_{F}}\sqrt{\mu_{\uparrow,\downarrow}^2-\Delta^2}$ is the Fermi momentum of the spin-$\uparrow$($\downarrow$) moving electron. We integrate in $p$, observing that the renormalization is the same as before, to obtain~\cite{Caldas2}:

\begin{eqnarray}
V_{eff}(\Delta,\mu_{\uparrow,\downarrow})=V_{eff}(\Delta) + \Theta_1 {\cal F}_{1} + \Theta_2 {\cal F}_{2},
\label{potefff}
\end{eqnarray}
where $V_{eff}(\Delta)$ is given by Eq.~(\ref{poteff2_1}), $\Theta_{1,2}=\Theta(\mu_{\uparrow,\downarrow}^2-\Delta^2)$ is the step function, defined as $\Theta(x)=0$, for $x\leq 0$, and $\Theta(x)=1$, for $x>0$, and ${\cal F}_{1,2} \equiv \frac{1}{2\pi \hbar v_F } \left[ \Delta^2 \ln \left(\frac{\mu_{\uparrow,\downarrow} + \sqrt{\mu_{\uparrow,\downarrow}^2-\Delta^2}}{\Delta} \right) -\mu_{\uparrow,\downarrow} \sqrt{\mu_{\uparrow,\downarrow}^2-\Delta^2} \right]$.

Although finite chemical potential is being considered now, the effective potential is unaffected if both $\mu_{\uparrow}$ and $\mu_{\downarrow}$ are smaller than $\Delta$, in which case it is reduced to that of Eq.~(\ref{poteff2_1}). For $\mu_{\uparrow,\downarrow}>\Delta$ the ground state is determined by jointly finding $\Delta_{0}(\mu_{\uparrow,\downarrow})$, and the analyzing of the effective potential at the minimum, $V_{eff}(\Delta_{0}(\mu_{\uparrow,\downarrow}))$.  Extremizing $V_{eff}(\Delta,\mu_{\uparrow,\downarrow})$ with respect to $\Delta$ yields the trivial solution ($\Delta=0$), and

\begin{equation}
\label{min2}
\ln \left( \frac{\Delta}{\Delta_0} \right) +  \frac{\Theta_1}{2} {\cal G}_1 + \frac{\Theta_2}{2} {\cal G}_2=0,
\end{equation}
where ${\cal G}_{1,2}\equiv \ln \left(\frac{\mu_{\uparrow,\downarrow} + \sqrt{\mu_{\uparrow,\downarrow}^2-\Delta^2}}{\Delta} \right)$, and we have made use of Eq.~(\ref{delta0}) to eliminate $m_F$ in the gap equation above. Equation~(\ref{min2}) can not be solved in a closed form for $\Delta$ as a function of $\mu_{\uparrow,\downarrow}$. Nevertheless, for $B_0=0$ i.e., for $\delta \mu=0$ which implies $\mu_{\uparrow}=\mu_{\downarrow}=\mu$, it can be easily solved:

\begin{equation}
\Delta_0 (\mu)= \sqrt{ \Delta_0(2 \mu - \Delta_0)}\;,
\label{Delta1}
\end{equation}
where $\Delta_0$ is given by Eq.~(\ref{delta0}), and $\frac{\Delta_0}{2} \leq \mu \leq \Delta_0$. However, the analysis of $V_{eff}(\Delta_0(\mu))$ shows that $\Delta_0 (\mu)$ is a local maximum and the solution $\Delta=\Delta_0$ represents a global minimum of $V_{eff}(\Delta,\mu)$ while $\mu < \mu_c$, (where the critical chemical potential, $\mu_c=\frac{\Delta_0}{\sqrt{2}}$, is obtained by $V_{eff}(\Delta=0,\mu_{c})=V_{eff}(\Delta=\Delta_0,\mu_c)$), with $\Delta_0=0$ representing a false minimum. When $\mu \geq \mu_c$, $\Delta(\mu \geq \mu_c)$ represents a local maximum and $\Delta_0=0$ is turned into a minimum through a first order phase transition~\cite{Wolff,CM}, agreeing with experiment~\cite{Fernando}. Thus, $\Delta_0$ as a function of the chemical potential in the absence of $B_0$ has the following expression:

\begin{equation}
\Delta_0(\mu)=\Theta(\mu_c-\mu) \Delta_0.
\label{gapmu}
\end{equation}

The situation changes considerably when $B_0$ is turned on, where $\Delta=0$ still represents the minimum of $V_{eff}(\Delta,\mu_{\uparrow,\downarrow})$ for $\delta \mu < \delta \mu_c = 0.38 \Delta_0$, where $\delta \mu_c$ is obtained from the equality $V_{eff}(\Delta=0,\mu_{\uparrow,\downarrow}(\delta \mu_c))=V_{eff}(\Delta=\Delta_0(\delta \mu_c),\mu_{\uparrow,\downarrow}(\delta \mu_c))$, with $\Delta_0(\delta \mu_c)$ being the solution of Eq.~(\ref{min2}) for $\mu_{\uparrow,\downarrow}(\delta \mu_c)$. For $\delta \mu \geq \delta \mu_c$, the $\Theta_{2}$ function prevents the ``$\downarrow$" term in the effective potential of participating in the minimum. In this case, Eq.~(\ref{min2}) can be rewritten as

\begin{equation}
\label{min4}
\Delta^4-2\mu_\uparrow \Delta_0^2 \Delta + \Delta_0^4=0.
\end{equation}

\subsubsection{Magnetic Properties at Zero Temperature}

As mentioned before, the application of a static magnetic field on the system results in a Zeeman energy given by $\Delta E= S_z g \mu_B B_0$, where $S_z=\pm 1/2$, $g$ ($\approx 2$ for {\it trans}-$\rm (CH)_x$) is the effective $g-$factor, $\mu_B=e \hbar/2m \approx  5.788 \times 10^{-5} ~{\rm eV}~ {\rm T}^{-1}$ is the Bohr magneton, $m$ is the bare electron mass, and $B_0$ is the magnetic field strength. Then we have $\delta \mu =|\Delta E|=\frac{g}{2}\mu_B B_0$. 

The number densities $n_{\uparrow,\downarrow}= -\frac{\partial}{\partial \mu_{\uparrow,\downarrow}} V_{eff}(\Delta,\mu_{\uparrow,\downarrow})$ are obviously imbalanced due to the asymmetry between $\mu_\uparrow$ and $\mu_\downarrow$, and will depend on $\delta \mu$. Before the critical asymmetry doping, $\Delta_0(\delta \mu)=0$ and the densities read

\begin{equation}
n_{\uparrow,\downarrow}(\delta \mu/\Delta_0 < 0.38)= \frac{1}{\pi \hbar v_F } \mu_{\uparrow,\downarrow}.
\label{nd}
\end{equation}
For such a low imbalance, compared to the critical chemical potential asymmetry $\delta \mu_c$, the total number density is the same as in the symmetric limit or, in other words, is independent of the applied magnetic field:

\begin{equation}
n_T=n_{\uparrow}+n_{\downarrow}=2n=\frac{2}{\pi \hbar v_F } \mu_c. 
\label{Nd1}
\end{equation}
In spite of that, there is a partial spin polarization

\begin{equation}
\delta n=n_{\uparrow}- n_{\downarrow}=\frac{2}{\pi \hbar v_F } \delta \mu = \frac{2}{\pi \hbar v_F } \mu_B B_0.
\label{Nd2}
\end{equation}

The zero $T$ Pauli magnetization of the polarized chain is defined as:

\begin{eqnarray}
M = \mu_B \delta n= \frac{2 \mu_B^2}{ \pi \hbar v_F} B_0.
\label{mag1}
\end{eqnarray}

The magnetic susceptibility is given by

\begin{equation}
\chi= \frac{\partial M}{\partial B_0}= \frac{2 \mu_B^2}{ \pi \hbar v_F},
\label{ms0}
\end{equation}
which is the well known Pauli expression of the magnetic susceptibility for noninteracting electrons at zero temperature.

\begin{figure}[htb]
  \vspace{0.5cm}
\epsfysize=5.5cm
\epsfig{figure=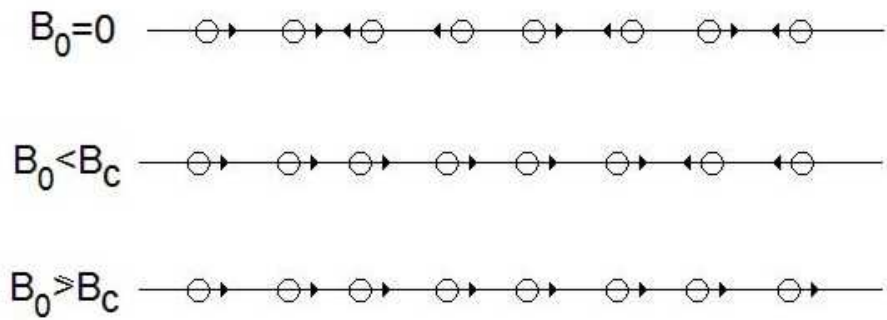,angle=0,width=12.2cm}
\caption[]{\label{Cartoon} Schematic representation of the spin orientations in the three possible situations in respect to the actuation of a Zeeman external constant magnetic field parallel to a $\rm (CH)_x$ wire. In the first line $B=0$ and consequently $M=0$. In the second line $B<B_{0,c}=2\delta \mu_c / g \mu_B$ implying in a net magnetization of the wire. Finally, when $B \geq B_{0,c}$ the wire is fully magnetized.}
\end{figure}

Increasing the asymmetry beyond the critical value $\delta \mu_c = \frac{g}{2}\mu_B B_{0, c}$, we find

\begin{equation}
n_{\uparrow}(\delta \mu/\Delta_0 > 0.38)= \frac{1}{\pi \hbar v_F }\sqrt{\mu_{\uparrow}^2-\Delta^2},
\label{Nd3}
\end{equation}
where $\Delta$ in the equation above is the solution of Eq.~(\ref{min4}) for a given $\delta \mu > \delta \mu_c$, and

\begin{equation}
n_{\downarrow}(\delta \mu/\Delta_0 > 0.38)= 0,
\label{Nd4}
\end{equation}
meaning that, effectively, there are only spin-$\uparrow$ electrons in the system i.e., the chain is fully polarized. The magnetization is now

\begin{eqnarray}
M_{f p} = \mu_B n_{\uparrow}= \frac{ \mu_B}{ \pi \hbar v_F} \sqrt{\mu_{\uparrow}^2-\Delta^2},
\label{magfully}
\end{eqnarray}
which also depends on the external magnetic field $B_0$ through $\Delta$. The results for the zero temperature magnetization of a $\rm (CH)_x$ wire under $B_0$ are displayed in Fig.~\ref{Cartoon}.

\subsubsection{Finite Temperature and Chemical Potentials}

The effects of finite temperature and density have been taken into account in the symmetric GN model ($\delta \mu =0$) by several authors. See, for instance, refs.~\cite{Wolff,Treml,Klimenko} and references in there.

We can rewrite $V_{eff}(\Delta,\mu_{\uparrow,\downarrow},T)$ as

\begin{equation}
V_{eff}(\Delta,\mu_{\uparrow,\downarrow},T)= V_{eff}(\Delta) + V_{eff}(\mu_{\uparrow,\downarrow},T),
\label{poteff3}
\end{equation}
where

\begin{eqnarray}
V_{eff}(\mu_{\uparrow,\downarrow},T)=
- k_B T \int^{\infty}_{0} \frac{dp}{\pi \hbar } ~ \Big[  \ln \left(1+e^{-\beta E_\uparrow^+}\right)+ \ln \left(1+e^{-\beta E_\uparrow^-}\right)
+ \ln \left(1+e^{-\beta E_\downarrow^+}\right) + \ln \left(1+e^{-\beta E_\downarrow^-}\right) \Big].
\label{poteff4}
\end{eqnarray}
Since we can not calculate expression (\ref{poteff4}) in a closed form, we shall use a high temperature expansion to evaluate it. Using the function

\begin{equation}
I(a,b)=\int_0^{\infty} dx \left[ \ln \left(1+ e^{-\sqrt{x^2+a^2}-b} \right) + \ln \left(1+ e^{-\sqrt{x^2+a^2}+ b} \right) \right],
\label{int1}
\end{equation}
where $a=\Delta/k_BT$, and $b=\mu/k_BT$, which can be expanded in the high temperature limit, $a<<1$ and $b<<1$, yielding, up to order $a^4$ and $b^2$~\cite{Rudnei},

\begin{equation}
I(a<<1,b<<1)= \frac{\pi^2}{6}+\frac{b^2}{2}-\frac{a^2}{2}\ln \left(\frac{\pi}{a} \right) -\frac{a^2}{4}(1-\gamma_E)-\frac{7 \xi(3)}{8 \pi^2} a^2 \left(b^2+\frac{a^2}{4}  \right) + \frac{186~\xi(5)}{128 \pi^4}  b^2 a^4 + {\cal O}\left(a^2 b^4 \right),
\label{int2}
\end{equation}
where $\gamma_E \approx 0.577... $ is the Euler constant and $\xi(n)$ is the Riemann zeta function, having the values $\xi(3) \approx 1.202$, and $\xi(5) \approx 1.037$. With the equation above, together with Eq.~(\ref{poteff2}), the high temperature asymmetrical effective potential is written as

\begin{eqnarray}
V_{eff}(\Delta,\mu_{\uparrow,\downarrow},T) &\equiv& V_{eff} = \frac{\Delta^2}{\pi \hbar v_F} \left[ \ln \left( \frac{\pi k_B T}{\Delta_0} \right) - \gamma_E \right]
-\frac{\pi}{3 \hbar v_F} (k_B T)^2 \\
\nonumber
&-& \frac{1}{2 \pi \hbar v_F} \left[ \mu_{\uparrow}^2+\mu_{\downarrow}^2 
- \frac{7 \xi(3)}{8 \pi^2} \frac{\Delta^4}{(k_B T)^2} - \frac{7 \xi(3)}{4 \pi^2} (\mu_{\uparrow}^2+\mu_{\downarrow}^2) \frac{\Delta^2}{(k_B T)^2} +
\frac{186 \xi(5)}{64 \pi^4} (\mu_{\uparrow}^2+\mu_{\downarrow}^2) \frac{\Delta^4}{(k_B T)^4} \right].
\label{poteff5}
\end{eqnarray}
The equation above may be rearranged in the form of a Ginzburg-Landau (GL) expansion of the grand potential density, which is appropriate to the analysis of the phase diagram in the region near the tricritical point,

\begin{equation}
\label{Veff}
V_{eff}= \alpha_0 + \alpha_2 \Delta^2 + \alpha_4 \Delta^4,
\end{equation}
where

\begin{eqnarray}
\label{coef}
\alpha_0(\mu_{\uparrow,\downarrow},T) &=& - \frac{1}{2 \pi \hbar v_F} \left[ \mu_{\uparrow}^2+\mu_{\downarrow}^2 \right] -\frac{\pi}{3 \hbar v_F} (k_B T)^2, \\
\nonumber
\alpha_2(\mu_{\uparrow,\downarrow},T) &=& \frac{1}{\pi \hbar v_F} \left[ \ln \left( \frac{\pi k_B T}{e^{\gamma_E} \Delta_0} \right) + \frac{7 \xi(3)}{8 \pi^2} \frac{(\mu_{\uparrow}^2+\mu_{\downarrow}^2)}{(k_B T)^2} \right], \\
\alpha_4(\mu_{\uparrow,\downarrow},T) &=& - \frac{1}{32 \pi^3 \hbar v_F (k_B T)^2} \left[ -14 \xi(3) + \frac{186 \xi(5)}{4 \pi^2} \frac{(\mu_{\uparrow}^2+\mu_{\downarrow}^2)}{(k_B T)^2} \right].
\end{eqnarray}
Extremizing $V_{eff}$ we find the trivial solution ($\Delta=0$) and the chemical potential and temperature dependent gap equation

\begin{equation}
\label{poteff6}
\Delta_0(\mu_{\uparrow,\downarrow},T)^2= -\frac{\alpha_2}{2 \alpha_4},
\end{equation}
which has meaning only if the ratio $\frac{\alpha_2}{ \alpha_4}$ is negative. Besides, a stable configuration (i.e., bounded from below) requires, up to this order, $\alpha_4>0$. At the minimum $V_{eff}$ reads

\begin{equation}
\label{Veff2}
V_{eff, min}= \alpha_0 - \frac{\alpha_2^2}{4 \alpha_4}.
\end{equation}

\begin{figure}[htb]
  \vspace{0.5cm}
\epsfysize=5.5cm
\epsfig{figure=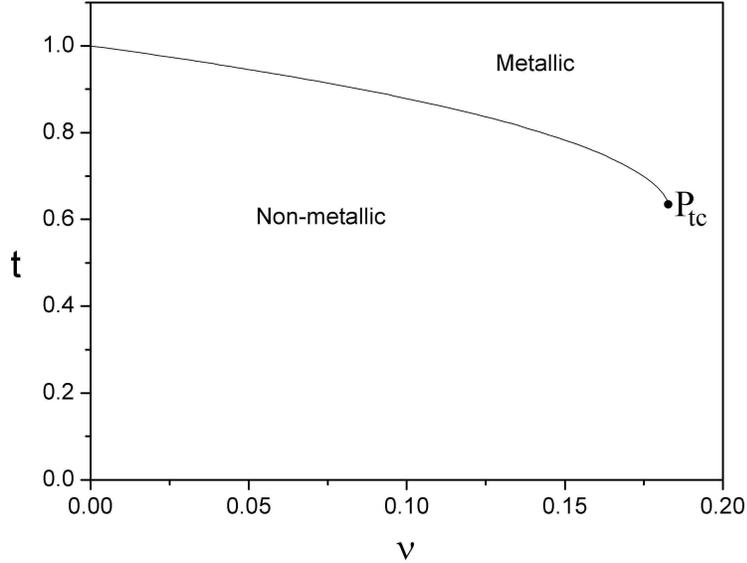,angle=0,width=12.2cm}
\caption[]{\label{pd1} The phase diagram $t=\frac{T}{T_c(0)}$ as a function of $\nu=\frac{7 \xi(3)}{8 \pi^2} \frac{(\mu_{\uparrow}^2+\mu_{\downarrow}^2)}{(k_B T_c(0))^2}$, from Eq.~(\ref{Tc1}). The second order transition line ends at the tricritical point $P_{tc}=(\frac{1}{2e},\frac{1}{\sqrt{e}})$. Below this point the transition is of first order.}
\end{figure}

The critical temperature $T_c$ is the temperature at which the coefficient of $\Delta^2$ vanishes i.e. $\alpha_2=0$, or

\begin{equation}
 \ln \left( \frac{\pi k_B T_c}{e^{\gamma_E} \Delta_0} \right) + \frac{7 \xi(3)}{8 \pi^2} \frac{(\mu_{\uparrow}^2+\mu_{\downarrow}^2)}{(k_B T_c)^2} =0,
\label{Tc1}
\end{equation}
where $T_c=T_c(\mu_{\uparrow},\mu_{\downarrow})$. As will become clear below, the equation above defines a second-order transition line separating the non-metallic ($\Delta \neq 0$) and metallic phases ($\Delta = 0$). At $\mu_{\uparrow}=\mu_{\downarrow}=0$, we recover the well-known result for the temperature at which the discrete chiral symmetry is restored~\cite{Jacobs}:

\begin{equation}
T_c(\mu_{\uparrow}=\mu_{\downarrow}=0) \equiv T_c(0)=\frac{e^{\gamma_E}}{\pi} \frac{\Delta_0}{k_B}.
\label{Tc2}
\end{equation}
In order to find $T_c(\mu_{\uparrow},\mu_{\downarrow})$ we define dimensionless variables $\nu =\frac{7 \xi(3)}{8 \pi^2} \frac{(\mu_{\uparrow}^2+\mu_{\downarrow}^2)}{(k_B T_c(0))^2}$ and $t=\frac{T}{T_c(0)}$, and with the help of Eq.~(\ref{Tc2}) we rewrite the L.H.S. of Eq.(\ref{Tc1}) as

\begin{equation}
y(t)=\ln(t) + \frac{\nu}{t^2}.
\label{Tc3}
\end{equation}
The zeros of $y(t)$ for a given $\nu$, i.e., for a given $\mu_{\uparrow}^2 + \mu_{\downarrow}^2$, are the respective $T_c$. This defines the (second-order) $T_c$ versus $\mu_{\uparrow}^2 + \mu_{\downarrow}^2$ phase diagram. A graphical inspection of $y(t)$~\cite{jstat2} shows that there is no solution for this function for $\nu$ above certain value, that we define $\nu_{tc}$. Besides, at $\nu_{tc}$ we have $y=y'=0$. These two equations give $t_{tc}$ and $\nu_{tc}$ for the tricritical point $P_{tc}=(\nu_{tc},t_{tc})$. Strictly speaking, $y$ and $y'$ are associated with the coefficients of the second-order and forth-order terms of the effective potential expanded in powers of $\Delta$~\cite{jstat1}. Solving the equations $y=0$ and $y'=0$ self-consistently (which is equivalent to solve $\alpha_2=\alpha_4=0$), we obtain the tricritical point analytically:

\begin{equation}
\nu_{tc}=\frac{7 \xi(3)}{8 \pi^2} \frac{(\mu_{\uparrow}^2 + \mu_{\downarrow}^2)_{tc}}{(k_B T_c(0))^2}=\frac{1}{2e},~~~~~~~t_{tc}=\frac{T_{tc}}{T_c(0)}=\sqrt{2\nu_{tc}}=\frac{1}{\sqrt{e}}.
\label{Tc4}
\end{equation}
For $\nu$ above certain value and less than $\nu_{tc}$, the function $y$ presents two solutions for $T_c$. However, the lower of these always corresponds to unstable solutions. The second-order transition curve, defined as the line starting at the point $(0,T_c(0))$ and ending at the point $((\mu_{\uparrow}^2+\mu_{\downarrow}^2)_{tc},T_{tc})$, comes simply from the solution of the gap equation. This curve, shown in Fig.~\ref{pd1}, represents a system at finite temperature where the chemical potentials $\mu_{\uparrow}^2+\mu_{\downarrow}^2=2(\bar\mu^2+{\delta \mu}^2)$ start from zero and increases until $(\mu_{\uparrow}^2+\mu_{\downarrow}^2)_{tc}$. Note that $\bar\mu^2+{\delta \mu}^2$ is zero if and only if $\bar\mu^2$ and ${\delta \mu}^2$ are both zero. It is well known that below the tricritical point one has to properly minimize the effective potential rather than using the gap equation as the transition becomes first order. Thus Eq.~(\ref{Tc1}) cannot be used for finding $T_c$ below $P_{tc}$ since this equation is valid only for the second-order transition. In this case $T_c$ has to be find numerically, through the equality $V_{eff}(T_c,\mu_{\uparrow},\mu_{\downarrow},\Delta=\Delta_{min})=V_{eff}(T_c,\mu_{\uparrow},\mu_{\downarrow},\Delta=0)$, where $\Delta_{min}$ is the non-trivial minimum of $V_{eff}$.

The number densities $n_{\uparrow,\downarrow}= -\frac{\partial}{\partial \mu_{\uparrow,\downarrow}} V_{eff}(T,\mu_{\uparrow,\downarrow},\Delta)$ read

\begin{equation}
n_{\uparrow,\downarrow}= \int^{\infty}_{0}{\frac{dp}{\pi \hbar}}~ \left[ n_k(E_{\uparrow,\downarrow}^{-}) - n_k(E_{\uparrow,\downarrow}^{+}) \right],
\label{n1}
\end{equation}
where $n_k(E_{\uparrow,\downarrow}^{+,-})=\frac{1}{e^{\beta E_{\uparrow,\downarrow}^{+,-}}+1}$ is the Fermi distribution function. The density difference

\begin{equation}
\delta n = n_{\uparrow}-n_{\downarrow}
\label{n2}
\end{equation}
is zero if $\delta \mu = \frac{g}{2} \mu_B B_0=0$, at any temperature, since in this case we have the equalities $n_k(E_{\uparrow}^{+})=n_k(E_{\downarrow}^{+})$, and $n_k(E_{\uparrow}^{-})=n_k(E_{\downarrow}^{-})$. The physical meaning of these results is that at zero external Zeeman magnetic field, the $\uparrow$ (up) and $\downarrow$ (down) electrons of the conduction $(+)$ band have the same density, and the same for the electrons of the valence $(-)$ band.

In the high temperature regime, the number densities are given by 

\begin{equation}
n_{\uparrow,\downarrow}(T)= \frac{1}{ \pi \hbar v_F} \left[ 1 - \frac{7 \xi(3)}{4 \pi^2} \frac{\Delta^2}{(k_B T)^2} \right]\mu_{\uparrow,\downarrow}.
\label{nd1}
\end{equation}
In the high temperature limit the total number density, $n_{T}(T)=n_{\uparrow}(T)+n_{\downarrow}(T)$, is independent of the applied field, as happens at zero $T$

\begin{equation}
n_{T}(T)= \frac{2}{ \pi \hbar v_F} \left[ 1 - \frac{7 \xi(3)}{4 \pi^2} \frac{\Delta^2}{(k_B T)^2} \right]\bar \mu,
\label{nT}
\end{equation}
and for the density difference we obtain

\begin{equation}
\delta n_{high~T}(T, \delta \mu)= \frac{2}{ \pi \hbar v_F} \left[ 1 - \frac{7 \xi(3)}{4 \pi^2} \frac{\Delta^2}{(k_B T)^2} \right]\delta \mu,
\label{nd2}
\end{equation}
that, as we have observed before, is clearly zero if $B_0=\delta \mu = 0$. Since the densities have to be evaluated at the minimum of the effective potential\footnote{Remembering that the thermodynamical potential per volume is defined as the free energy density (or effective potential in the field theory language we are using here) at its minimum, $\Omega(T,\mu_{\uparrow,\downarrow}) = V_{eff}(T,\mu_{\uparrow,\downarrow},\Delta_{min})$.}, we use Eq.~(\ref{poteff6}) in the equation above and find the temperatures at which the densities and, consequently, the density difference vanish. These temperatures are the solutions of

\begin{equation}
\gamma_E -\frac{1}{2} + \ln \left( \frac{\Delta_0}{\pi k_B T_c^*} \right) - \frac{7 \xi(3)}{8 \pi^2} \frac{(\mu_{\uparrow}^2+\mu_{\downarrow}^2)}{(k_B T_c^*)^2} =0,
\label{nd3}
\end{equation}
where $T_c^*=T_c^*(\mu_{\uparrow},\mu_{\downarrow})$. As for the gap parameter, the equation above defines the second-order line for the densities and the density imbalance. At $\mu_{\uparrow}=\mu_{\downarrow}=0$, we get

\begin{equation}
T_c^*(\mu_{\uparrow}=\mu_{\downarrow}=0) \equiv T_c^*(0)=\frac{e^{\gamma_E-\frac{1}{2}}}{\pi} \frac{\Delta_0}{k_B}.
\label{Tc*1}
\end{equation}
It is very easy to see that $T_c^*(0)=\frac{T_c(0)}{\sqrt{e}}$, which coincides with $T_{tc}$, where $T_{tc}$ is given by Eq.~(\ref{Tc4}). Proceeding as before we find

\begin{equation}
\nu_{tc}^*=\frac{7 \xi(3)}{8 \pi^2} \frac{(\mu_{\uparrow}^2+\mu_{\downarrow}^2)_{tc}}{(k_B T_c^*(0))^2}=\frac{1}{2e^2},~~~~~~~t_{tc}^*=\frac{T_{tc}^*}{T_c^*(0)}=\sqrt{2\nu_{tc}^*}=\frac{1}{e},
\label{nd4}
\end{equation}
defining the tricritical point for the densities, total density and density imbalance second order curves.

Let us now verify the possibility of a fully polarized state at finite temperature. It would be possible with a magnetic field with a intensity such that $n_{\downarrow}$ in Eq.~(\ref{nd1}) vanishes. In this case $\mu_{\downarrow}=\bar \mu-\delta \mu_c=\mu_c- \frac{g}{2} \mu_B B_{0,c}=0$, or

\begin{equation}
B_{0,c}= \frac{2 \mu_c}{g \mu_B}.
\label{cmf}
\end{equation}

\subsubsection{Magnetic Properties at High Temperature}

The Pauli magnetization of the chain in the high temperature limit has the following expression:

\begin{eqnarray}
M_{high~T}(T) &=& \mu_B \delta n_{high~T}(T)= \frac{2 \mu_B}{ \pi \hbar v_F} \left[ 1 - \frac{7 \xi(3)}{4 \pi^2} \frac{\Delta^2}{(k_B T)^2} \right]\delta \mu \nonumber \\
&=& \frac{g \mu_B^2}{ \pi \hbar v_F}\left\{1- 2\left[\ln \left( \frac{T}{T_c(0)} \right) + \frac{7 \xi(3)}{8 \pi^2} \frac{(\mu_{\uparrow}^2+\mu_{\downarrow}^2)}{(k_B T)^2} \right] \right\} B_0,
\label{mag2}
\end{eqnarray}
where $T_c(0)$ is given by Eq.~(\ref{Tc2}) and we have made use of Eq.~(\ref{poteff6}) to leading order in $\frac{(\mu_{\uparrow}^2+\mu_{\downarrow}^2)}{(k_B T)^2}$. The second-order line where the magnetization vanishes is the same as the one given by Eq.~(\ref{nd3}). With $M_{high~T}(T)$ we are in condition to obtain the magnetic susceptibility in this regime

\begin{equation}
\chi_{high~T}(T)= \frac{\partial M_{high~T}(T)}{\partial B_0}= \chi(0)- \chi(T),
\label{ms1}
\end{equation}
where

\begin{equation}
\chi(0)=\frac{g \mu_B^2}{ \pi \hbar v_F},
\label{ms2}
\end{equation}
is the zero temperature Pauli magnetic susceptibility for noninteracting electrons given by Eq.~(\ref{ms0}), and

\begin{equation}
\chi(T)=\frac{2 g \mu_B^2}{ \pi \hbar v_F} \left[\ln \left(\frac{T}{T_c(0)} \right) + \frac{7 \xi(3)}{4 \pi^2(k_B T)^2} \left( \mu_c^2 + \frac{3}{4}g^2 \mu_B^2 B_0^2 \right) \right].
\label{ms3}
\end{equation}
The function $\chi_{high~T}(T)$ also behaves at finite temperature as the densities and the magnetization, with a second-order transition up to a tricritical point given by Eq.~(\ref{nd4}). Below this point the transition is again of first order. 

Note that despite the fully polarization, at $B_{0,c}$ the magnetization is given by exactly the same expression shown in Eq.~(\ref{mag2}). With the help of Eq.~(\ref{nd1}) we find:

\begin{eqnarray}
M_{high~T,c}(T) = \mu_B n_{\uparrow}(T) &=&  \frac{\mu_B}{ \pi \hbar v_F} \left[ 1 - \frac{7 \xi(3)}{4 \pi^2} \frac{\Delta^2}{(k_B T)^2} \right](\mu_c + \delta \mu_c) \nonumber \\
&=& \frac{ g \mu_B^2}{ \pi \hbar v_F}\left\{1-2 \left[ \ln \left( \frac{T}{T_c^*(0)} \right) + \frac{7 \xi(3)}{8 \pi^2} \frac{(\mu_{\uparrow}^2+\mu_{\downarrow}^2)}{(k_B T)^2} \right]\right\} B_{0,c}.
\label{mag3}
\end{eqnarray}
This shows that this function is indeed continuous for $0\leq B_0 \leq B_{0,c}$.

\subsection{Inhomogeneous $\Delta(x)$ Condensates}

Since we consider the addition of a chemical potential (i.e., doping) in the theory representing a Peierls distorted material like TPA, and the effects of a Zeeman magnetic field on these materials, some important remarks are in order. It is well-known that doping in conducting polymers with degenerate ground states results in lattice deformation, or non-linear excitations, such as kink solitons and polarons, meaning that $\Delta(x)$ can vary in space~\cite{Horovitz1,Horovitz2,Review}. Therefore, one may expect not only homogeneous-like configurations (as considered in the previous subsection), but also that the inclusion of these excitations in any theoretical calculation in this model should be considered. In this context, within the GN field theory model that we are considering, by taking into account kink-like configurations in the large $N$ approximation, the authors of Refs.~\cite{gnpolymers,gnpolymers2,gnpolymers3,Basar} found evidence for a crystalline phase that shows up in the extreme $T \sim 0$ and large $\mu$ part of the phase diagram, while the other extreme of the phase diagram, for large $T$ and small $\mu$, seemed to remain identical to the usual large $N$ results for the critical temperature and tricritical points, which are well-known results \cite{Wolff} for the GN model. The restoration of the broken discrete chiral symmetry of the massless GN model, which is signalized when the kinkantikink crystalline condensate~\cite{Saxena,gnpolymers2,Schnetz} transforms into the kink crystal, is explained when an exotic supersymmetric structure is properly considered~\cite{Mikhail}.

To take into account the effects of inhomogeneous configurations in the GL expansion of the grand potential density, let us write Eq.~(\ref{Veff}) in terms of $\Delta(x)$ and its derivatives up to $\alpha_4$~\cite{gnpolymers,gnpolymers2,gnpolymers3,Basar}:

\begin{equation}
\label{Veff(x)}
V_{eff}(x)= \alpha_0 + \alpha_2 \Delta(x)^2 + \alpha_4 [ \Delta(x)^4 + \Delta(x)'^2 ],
\end{equation}
where $\Delta(x)' \equiv d \Delta(x) / dx$. A straightforward variational calculation gives the following condition for the minimization of the free energy $E = \int V_{eff}(x)~dx$:

\begin{equation}
\label{difeq1}
\Delta(x)'' -2 \Delta(x)^3 - \frac{\alpha_2}{\alpha_4}\Delta(x)=0.
\end{equation}
The general solution of an equation of the form

\begin{equation}
\label{difeq2}
\Delta(x)'' -2 \Delta(x)^3 + (1+ \varsigma ) \Delta_0^2 \Delta(x)=0,
\end{equation}
can be written as~\cite{Basar}

\begin{equation}
\label{difeq3}
\Delta(x,\varsigma) \equiv \Delta(x) = \Delta_0 \sqrt{\varsigma} ~{\rm sn}(\Delta_0 x; \varsigma ),
\end{equation}
where ${\rm sn}$ is the Jacobi elliptic function with the real elliptic parameter $0 \leq \varsigma   \leq  1$. The ${\rm sn}$ function has period $2 {\bf K}(\varsigma)$, where ${\bf K}(\varsigma  ) \equiv \int_0^{\pi/2}[1-\varsigma   \sin^2(z)]^{-1/2} dz$ is the complete elliptic integral of first kind. At $T=0$ and finite density (or $\mu$), the $\varsigma$ parameter can be determined by a variational procedure yielding an average fermion density dependent transcendental equation for $\varsigma$, which can be solved in the low- and high-density limits~\cite{gnpolymers3}. At finite $T$ and $\mu$, the $\varsigma$ parameter can be determined within the Schnetz, Thies and Urlichs' ansatz to solve the inhomogeneous gap parameter problem, together with the scale factor $A$ by a numerical minimization of the renormalized grand canonical potential~\cite{gnpolymers}. 

$\Delta(x)$ in (\ref{difeq3}) represents an array of real kinks. When $\varsigma  =1$ Eq.~(\ref{difeq3}) is reduced to the single kink condensate given in Eq.~(\ref{kink}). By a direct comparison between Eqs.~(\ref{difeq1}) and (\ref{difeq2}) one can easily identify the scale parameter $\Delta_0$ at high $T$ (i.e. $T$ near $T_c$) as

\begin{equation}
\label{difeq4}
\Delta_0=\sqrt{\left( -\frac{\alpha_2}{\alpha_4} \right) \frac{1}{1+\varsigma }} \equiv \Delta_0(\varsigma).
\end{equation}
Since $\frac{1}{1+\varsigma} > 0$, the solution for inhomogeneous condensates has physical meaning only if the ratio $\frac{\alpha_2}{\alpha_4}$ is negative, as in the case of homogeneous condensates.

In order to see the effects of Eq.~(\ref{difeq3}) in the grand potential density, we add the term $-\alpha_4 \frac{1}{3}(\Delta(x)^2)''$ to $V_{eff}(x)$ in Eq.~(\ref{Veff(x)}). It is easy to see that this term does not alter the solution for the minimum of the free energy, Eq.~(\ref{difeq1}), but allows the use of the following important identities supported by the solution in (\ref{difeq2}):

\begin{eqnarray}
\label{difeq5}
(\Delta(x)')^2 &=& \Delta^4(x) - (1+\varsigma) \Delta_0^2 \Delta(x)^2 + \varsigma  \Delta_0^4,\\
\nonumber
(\Delta(x)^2)^{''} &=& 6\Delta^4(x) - 4(1+\varsigma) \Delta_0^2 \Delta(x)^2 + 2 \varsigma  \Delta_0^4.
\end{eqnarray}
With the equations above, we can write the $x$-dependent grand potential density as

\begin{equation}
\label{Veff(x)2}
V_{eff}(x)= \alpha_0 + \alpha_2\Delta(x)^2 + \alpha_4 \frac{1}{3} \left[ (1+\varsigma)\Delta_0^2 \Delta(x)^2 + \varsigma  \Delta_0^4 \right].
\end{equation}
Averaging over one period, it is found~\cite{Basar} that $<\Delta(x)^2>= \left(1-\frac{{\bf E}(\varsigma)}{{\bf K}(\varsigma )} \right) \Delta_0^2 $, where ${\bf E}(\varsigma)$ is the complete elliptic integral of second kind. The ratio ${\bf E}(\varsigma)/{\bf K}(\varsigma)$ is a smooth function of $\varsigma$ interpolating monotonically between $0$ and $1$. Thus we can write

\begin{equation}
\label{Veff(x)3}
<V_{eff}(x)> = \alpha_0 +  {\cal A}_2 \Delta_0^2  + {\cal A}_4 \Delta_0^4,
\end{equation}
where 

\begin{eqnarray}
\label{Newcoef}
{\cal A}_2 &=& \alpha_2 \left(1-\frac{{\bf E}(\varsigma)}{{\bf K}(\varsigma)} \right),\\
\nonumber
{\cal A}_4 &=& \alpha_4 \frac{1}{3} \left[\varsigma + (1+ \varsigma)\left(1-\frac{{\bf E}(\varsigma)}{{\bf K}(\varsigma)} \right)\right].
\end{eqnarray}
\newline
As we will see next, with the consideration of inhomogeneous $\Delta(x)$ condensates we obtain very interesting results. This generalizes the analysis of the grand potential density, such that the results obtained in subsection {\bf A} for homogeneous condensates are particular cases which are obtained when specific values of the elliptic parameter $\varsigma$ are taken:
\newline
\newline 
(${\bf I.}$) For $\varsigma =0$, $\frac{{\bf E}(\varsigma =0)}{{\bf K}(\varsigma =0)}=1$, which implies ${\cal A}_2={\cal A}_4=0$, and the grand potential density is that of the metallic phase, for which $\Delta=0$ and $V_{eff}=\alpha_0$.
\newline
\newline
(${\bf II.}$) For $\varsigma =1$, $\frac{{\bf E}(\varsigma =1)}{{\bf K}(\varsigma=1)}=0$, so the grand potential density is that of the homogeneous case, Eq.~(\ref{Veff}), at the non-trivial minimum $\Delta_0$: $V_{eff}= \alpha_0 + \alpha_2 \Delta_0^2 + \alpha_4 \Delta_0^4$.
\newline
\newline
Besides, we find that:
\newline
\newline
(${\bf III.}$) For a $0< \varsigma  < 1$, the grand potential density is that of Eq.~(\ref{Veff(x)3}) that, as expected, is the interpolation of the two previous ones.
\newline
\newline
And finally, we find that the tricritical point is ``stationary" under the application of an external Zeeman magnetic field on $\rm CH$ chains:
\newline
\newline
(${\bf IV.}$) For any $0 <   \varsigma \leq  1$ it is easy to see from Eq.~(\ref{Veff(x)3}) that the tricritical point is still found for $\alpha_2=\alpha_4=0$. These coefficients are $\mu_{\uparrow,\downarrow}$ and $T$ dependent and were not affected by the space dependence of the condensate $\Delta(x)$, as seen in the symmetric case ($\delta \mu =B_0=0$)~\cite{gnpolymers,gnpolymers2,gnpolymers3,Basar}. The novelty is that the location of the tricritical point of the TPA phase diagram under the influence of an external Zeeman magnetic field is also unaltered even considering a $x$ dependent grand potential density. This happens because in the high temperature limit the influence of the Zeeman field is not sufficient to change the position of the tricritical point.

\subsection{The Gross-Neveu Model Beyond the Mean-Field Approximation}

We now investigate the GN model beyond the mean field or large N approximation employing the optimized perturbation theory (OPT). The OPT is a method to take into account non-perturbative contributions to the mean-field results, which are of crucial importance to recover the reliability of perturbation theory at finite temperature, mainly in theories displaying thermodynamical phase transitions~\cite{CaldasFT}. For a better description of the implementation of the OPT procedure, please see refs.~\cite{linear,Rudnei}. The OPT is closely related with the optimized expansion (OE) method. As OPT, the OE is an alternative non-perturbative technique which also allows one to go beyond the framework of the standard perturbative expansion~\cite{OE}. 

Applying the usual OPT interpolation prescription to the {\it original} Lagrangian density given by Eq.~(\ref{LagGN}), we define the interpolated theory:

\begin{equation}
{\cal L}_{\rm GN}^{\delta}(\psi, {\bar \psi}) =
\sum_s \bar{\psi}^s
\left( i \hbar \gamma_0 \partial_t - i \hbar v_F \gamma_1 \partial_x
\right ) \psi^s  -
\eta (1-\delta) {\bar \psi} \psi
+ \delta \frac {\lambda_{\rm GN}}{2N} \hbar v_F({\bar \psi} \psi)^2\;.
\label{GNlde}
\end{equation}

\noindent
It is easy to verify that at $\delta=0$ we have a theory of free fermions, and
the original theory is recovered for $\delta=1$. $\eta$ is an arbitrary mass
parameter. Now, by re-introducing the scalar field $\Delta$, which can be
achieved by adding the quadratic term corresponding to a Hubbard-Stratonovich
trick:

\begin{equation}
- \frac{ \delta N}{2 \hbar v_F \lambda_{\rm GN}} \left ( \Delta +
\frac {\lambda_{\rm GN}}{N} \hbar v_F {\bar \psi} \psi \right )^2 \,,
\end{equation}
to ${\cal L}_{\rm GN}^{\delta}(\psi, {\bar \psi})$, one obtains the interpolated model
corresponding to the original TLM model given by Eq.~(\ref{LagTLM}),

\begin{equation}
{\cal L}_{\rm GN}^{\delta} =
\sum_{s=1,2} \bar{\psi}^s
\left( i \hbar \gamma_0 \partial_t - i \hbar v_F \gamma_1 \partial_x
\right ) \psi^s -
\delta \Delta {\bar \psi} \psi - \eta (1-\delta) {\bar \psi} \psi
- \frac {\delta N }{2 \hbar v_F \lambda_{\rm GN} } \Delta^2   \;.
\label{GNdelta}
\end{equation}
Introducing the chemical potentials as before and considering the actuation of the Zeeman field, as explained below Eq.~(\ref{action1}), we write:

\begin{equation}
\label{L1Rew2}
{ L}_{\rm GN}^{\delta}= \sum_{s=1,2} \bar \psi^s [- \gamma_0 \hbar \partial_\tau + i \hbar v_F \gamma_1 \partial_x +  \gamma_0 \mu_s ] \psi^s -
\delta \Delta {\bar \psi} \psi - \eta (1-\delta) {\bar \psi} \psi
- \frac {\delta N }{2 \hbar v_F \lambda_{\rm GN} } \Delta^2.
\end{equation}

\noindent 
Since Eq.~(\ref{L1Rew2}) in the limit $\delta \mu \to 0$ is the same model already studied in Ref.
\cite{Rudnei}, so we do not repeat all the details related to the free energy
density derivation here, where only the main steps and results relevant for
our application to the {\rm CH}-chains will be presented.

Generally, the OPT method can be implemented as follows. Any physical
quantity, $F^{(k)}$, is {\it perturbatively} computed from the interpolated
model, up to some finite order-$k$ in $\delta$, which is formally used only as
a bookkeeping parameter which is set to the unity, at the end of calculation.
But in this process any (perturbative) result at order $k$ in the OPT remains
$\eta$ dependent.  This arbitrary (a priori) parameter is then fixed by a
variational method that then generates nonperturbative results, in the sense
that it resums to all orders a certain class of perturbative contributions
through self-consistent equations. Such optimization method is known as the
principle of minimal sensitivity (PMS) and amounts to require that
$F^{(k)}$ be evaluated at the point where it is less sensitive to this
parameter.  This criterion translates into the variational relation \cite{pms}

\begin{equation} 
\frac {d F^{(k)}}{d \eta}\Big |_{ \delta=1, \eta=\bar \eta} = 0 \;.
\label{PMS} 
\end{equation}

\noindent
The optimum value $\bar \eta$ that satisfies Eq. (\ref{PMS}) must be a
function of the original parameters, including the couplings, thus generating
``non-perturbative" results.  In our case, we are interested in evaluating the
optimized free energy at finite temperature and density for the scalar field,
$\Delta$, once the fermions have been integrated out.

\subsubsection{The Optimized Free Energy Density}

We generalize to the asymmetrical case the Landau's free energy density (or effective potential, in
the language of quantum field theories) to order-$\delta$, that has been evaluated in Ref. \cite{Rudnei} using usual functional and diagrammatic techniques. The result is:

\begin{eqnarray}
{\cal F} (\Delta, \eta, T, \mu_{\uparrow,\downarrow}) &=&
\delta \frac {N \Delta^2}{2 \lambda_{\rm GN} v_F \hbar} -
 \frac{N}{2\pi v_F \hbar } \left \{ \eta^2 \left [ \frac {1}{2} + \ln \left (
\frac {m_F}{\eta} \right ) \right ] +  (kT)^2 [I_1(\eta,\mu_\uparrow,T)+I_1(\eta,\mu_\downarrow,T)] \right \}
\nonumber \\
&+& \delta \frac{N\eta(\eta-\Delta)}{\pi v_F \hbar}
\left[\ln\left(\frac{m_F}{\eta}\right) - \frac{1}{2} \left(I_2 (\eta,\mu_\uparrow,T) + I_2 (\eta,\mu_\downarrow,T) \right) \right]
\nonumber \\
&+& \delta \frac { \lambda_{\rm GN}}{4\pi^2 v_F \hbar}  \eta^2
  \left [ \ln \left ( \frac {m_F}{\eta}  \right ) - \frac{1}{2} \left(I_2 (\eta,\mu_\uparrow,T) + I_2 (\eta,\mu_\downarrow,T) \right)\right ]^2 
\nonumber \\
&+& \delta \frac { \lambda_{\rm GN}}{4\pi^2 v_F \hbar}
\frac{1}{2}(kT)^2 \left[ I^2_3(\eta,\mu_\uparrow,T) + I^2_3(\eta,\mu_\downarrow,T) \right] \;.
\label{Vdelta1}
\end{eqnarray}
where the functions $I_1$, $I_2$ and $I_3$ are given respectively by

\begin{equation}
I_1(\eta,\mu,T) = \int_0^\infty dx
\left\{ \ln \left[ 1+
e^{-\sqrt{x^2+\left(\frac{\eta}{kT}\right)^2}-\frac{\mu}{kT} } \right]
+ \ln \left[ 1+e^{-\sqrt{x^2+\left(\frac{\eta}{kT}\right)^2}+\frac{\mu}{kT} } 
\right] \right\}\;,
\label{I1}
\end{equation}

\begin{equation}
I_2(\eta,\mu,T)=\int_0^\infty \frac {d x}{\sqrt{x^2 +\eta^2/(kT)^2}} \left[
\frac{1}{e^{\sqrt{x^2+\left(\frac{\eta}{kT}\right)^2}+\frac{\mu}{kT} }+1}+
\frac{1}{e^{ \sqrt{x^2+\left(\frac{\eta}{kT}\right)^2}-\frac{\mu}{kT}  }+1} 
\right] \;,
\label{I2}
\end{equation}
and

\begin{eqnarray}
I_3 (\eta,\mu,T)
= \sinh\left(\frac{\mu}{kT}\right) \int_0^\infty d x
\frac{1}{\cosh\left(\sqrt{x^2 + \frac{\eta^2}{(kT)^2} }\right) + 
\cosh\left(\frac{\mu}{kT}\right)}  \;.
\label{I3}
\end{eqnarray}
In Eq.~(\ref {Vdelta1}), $\Delta$ is the homogeneous (constant field) configuration for the
scalar field discussed already in the subsection~\ref{homogeneous}.  In the computation performed in Ref.~\cite{Rudnei}, the free energy density has been renormalized using the $\overline {\rm MS}$ scheme for dimensional regularization. To a direct comparison between our results and that of Ref.~\cite{Rudnei}, we have adopted dimensional regularization in this section instead of the cutoff regularization we employed in the previous section. As will be clear soon, Eq.~(\ref{Vdelta1}), evaluated at first order in the OPT, already takes into account corrections beyond the large-$N$ (or mean-field) leading order result.

By optimizing Eq.~(\ref{Vdelta1}) through the PMS condition, Eq. (\ref{PMS}), we obtain the optimum value for the parameter $\eta = \bar{\eta}$, which is then re-inserted back in Eq. (\ref{Vdelta1}), allowing us to compute the order
parameter ${\overline \Delta}$ that minimizes the free energy. {}Using the PMS procedure we then obtain, from Eq.~(\ref{Vdelta1}) at $\eta = {\bar \eta}$ and $\delta=1$, the general result that factors into:

\begin{eqnarray}
\left [ {\cal Y}(\eta,\mu_{\uparrow,\downarrow},T)
+ \eta \frac{d}{d \eta} {\cal Y}(\eta,\mu_{\uparrow,\downarrow},T) \right ]
\left[ \eta - \Delta +
\eta \frac{\lambda_{\rm GN}}{2 \pi N} {\cal Y}(\eta,\mu_{\uparrow,\downarrow},T) \right] \Bigr|_{\eta = \bar{\eta}}
\nonumber
\\
+ \frac{(kT)^2 \lambda_{\rm GN}}{4 \pi N} \left[I_3(\eta,\mu_{\uparrow},T)
\frac {d}{d \eta}I_3(\eta,\mu_{\uparrow},T) + I_3(\eta,\mu_{\downarrow},T)
\frac {d}{d \eta}I_3(\eta,\mu_{\downarrow},T)\right]
\Bigr|_{\eta = \bar{\eta}} = 0 \;,
\label{genpms}
\end{eqnarray}
where we have defined the function

\begin{equation}
\label{y}
{\cal Y}(\eta,\mu_{\uparrow,\downarrow},T) \equiv \ln \left ( \frac {m_F}{\eta}  \right ) - \frac{1}{2} \left(I_2 (\eta,\mu_\uparrow,T) + I_2 (\eta,\mu_\downarrow,T) \right) \;.
\end{equation}
Note that when $N \to \infty$ in Eq. (\ref{genpms}), one obtains the solution ${\bar \eta} = \Delta$ and the mean-field standard result is exactly reproduced, as it should be \cite{Rudnei,npb}.

\subsubsection{ The Gap Energy at Zero Temperature and Chemical Potential Beyond Mean-Field}

In order to perform a numerical analysis we must fix all parameters. This can be done by considering the gap energy. In the GN language the order parameter ${\overline \Delta}$ is just the TLM gap parameter which, at $T=0$ and
$\mu=0$, has been denoted as $\Delta_0$ in Eq.~(\ref{delta0}). At $T=\mu=0$ we have that $I_1=I_2=I_3=0$ so that Eq.~(\ref{genpms}) and $d{\cal F}/d \Delta=0$ give, at order-$\delta$, respectively:

\begin{equation}
\Delta= \bar \eta \left[ 1 +  \frac{\lambda_{\rm GN}}{2\pi N} \ln \left( \frac{m_F}{\bar \eta} \right) \right]\;,
\label{eta}
\end{equation}
and

\begin{equation}
\Delta= \frac{\lambda_{\rm GN}}{\pi} \bar \eta   \ln \left( \frac{m_F}{\bar \eta} \right) \;.
\label{delt}
\end{equation}
Plugging $\bar \eta   \ln \left( \frac{m_F}{\bar \eta} \right)=\frac{\pi\Delta}{\lambda_{\rm GN}}$ from Eq.~(\ref{delt}) in Eq.~(\ref{eta}) we find.

\begin{equation}
\Delta= \left ( 1- \frac{1}{2  N} \right )^{-1} \bar \eta \;.
\label{eta2}
\end{equation}
Solving these equations self-consistently yields:

\begin{equation}
 \bar \eta_{ \delta^1}(T=\mu_{\uparrow,\downarrow}=0) \equiv \bar \eta_{ \delta^1} (0) =
m_F   e^{-\frac {\pi}{\lambda_{\rm GN}\left ( 1- \frac{1}{2  N} \right ) }} \;,
\label{eta0}
\end{equation}
and

\begin{equation}
 \Delta_{ \delta^1}(T=\mu_{\uparrow,\downarrow}=0) \equiv \Delta_{ \delta^1}(0)=
F(N)^{-1} m_F  e^{
 -\frac {\pi}{\lambda_{\rm GN}F(N) }} \;,
\label{deltaopt}
\end{equation}
where $F(N)=\left( 1- \frac{1}{2  N} \right)$. As pointed out in Ref.~\cite{Rudnei}, Eq. (\ref{deltaopt}) explicitly includes corrections beyond large $N$, as obtained from the OPT approach, to the mean-field result. More precisely, taking the mean-field approximation, $N\to \infty$ in Eq.~(\ref{deltaopt}) and using the relation $\lambda_{\rm GN}=N \pi \lambda_{\rm TLM}$, the OPT result exactly recovers the mean field result for $N=2$ \cite{Review}, as expected~\cite{Rudnei,npb}. Besides, we have that $\Delta_{ \delta^1}(0)$ in the limit $N\to \infty$ reproduces Eq.~(\ref{deltaopt}) with an appropriate redefinition of the (arbitrary) renormalization scale $m_F$.

It is worth to mention that, usually, in a renormalizable quantum field theory, one can choose arbitrary value for $m_F$ and $\lambda_{\rm GN}$ will run with the scale appropriately, at a given perturbative order, so that $\Delta_0$ remains scale-invariant as dictated by the renormalization group~\cite{PRB}.

\subsubsection{ The Gap Energy at Zero Temperature and Finite Chemical Potential Beyond Mean-Field}

We start by taking the $T \to 0$ limit of the functions defined by Eqs.~(\ref{I1}),~(\ref{I2}) and~(\ref{I3}) which appear in the free energy density, Eq.~(\ref {Vdelta1}). In this limit, these equations read~\cite{Rudnei}:

\begin{eqnarray}
&& \lim_{T\to 0} T^2 I_1(\eta,\mu,T)=
- \frac{1}{2} \theta(\mu - \eta)
\left[ \eta^2 \ln \left( \frac{\mu + \sqrt{\mu^2 -
\eta^2}}{\eta}
\right) - \mu \sqrt{\mu^2 - \eta^2} \right]\;,
\label{I1T0}
\\
&&
\lim_{T \to 0} I_2(\eta,\mu,T) =  \theta(\mu-\eta)
\ln\left(\frac{\mu +\sqrt{\mu^2 - \eta^2}}{\eta} \right)\;,
\label{I2T0}
\\
&&
\lim_{T \to 0} T^2 I_3(\eta,\mu,T) =  {\rm sgn}(\mu) \theta(\mu-\eta)
\sqrt{\mu^2 - \eta^2} \;.
\label{I3T0}
\end{eqnarray}

Using Eqs.~(\ref{I1T0}), (\ref{I2T0}) and (\ref{I3T0}) in the PMS equation, Eq.~(\ref{genpms}), we notice that it can be divided into two cases: i) $\mu_{\uparrow,\downarrow} < \eta$, which is equivalent to the situation where $T=\mu_{\uparrow,\downarrow}=0$ studied in the previous subsection, and ii) $\mu_{\uparrow,\downarrow} > \eta$. For $\mu_{\uparrow,\downarrow} > \eta$ we have

\begin{equation}
[{\cal Y}(\eta,\mu_{\uparrow,\downarrow})+G(\eta,\mu_{\uparrow,\downarrow})][\eta - \Delta + \frac{\lambda_{GN}}{2 \pi N} \eta {\cal Y}(\eta,\mu_{\uparrow,\downarrow})] - \frac{\lambda_{GN}}{2 \pi N} \eta =0\;,
\label{PMS1}
\end{equation}
where

\begin{equation}
{\cal Y}(\eta,\mu_{\uparrow,\downarrow},T=0) \equiv {\cal Y}(\eta,\mu_{\uparrow,\downarrow})= \ln \left ( \frac {m_F}{\eta}  \right ) - \frac{1}{2} \left[ \ln \left(\frac{\mu_{\uparrow}+\sqrt{\mu_{\uparrow}^2 -\eta^2}}{\eta} \right) + \ln \left(\frac{\mu_{\downarrow}+\sqrt{\mu_{\downarrow}^2 -\eta^2}}{\eta} \right) \right]\;,
\label{Y1}
\end{equation}
and we have defined the function

\begin{equation}
G(\eta,\mu_{\uparrow,\downarrow}) \equiv G=\frac{1}{2} \left(\frac{\mu_{\uparrow}}{\sqrt{\mu_{\uparrow}^2 -\eta^2}}+ \frac{\mu_{\downarrow}}{\sqrt{\mu_{\downarrow}^2 -\eta^2}} \right)-1\;.
\label{G}
\end{equation}
Using 

\begin{equation}
{\cal Y}(\eta,\mu_{\uparrow,\downarrow})=\frac{ \pi \Delta}{\lambda_{GN} \bar\eta}\;,
\label{Y}
\end{equation}
from $d{\cal F}/d \Delta=0$ in Eq.~(\ref{PMS1}), we find a second order equation for $\Delta$:

\begin{equation}
\frac{ \pi }{\lambda_{GN} \bar \eta}\left(1-\frac{1}{2N} \right) \Delta^2 -\left[\frac{ \pi }{\lambda_{GN} } -G\left(1-\frac{1}{2N} \right) \right] \Delta - \left( G - \frac{\lambda_{GN}}{ 2 \pi N} \right)\bar \eta=0\;,
\label{PMS2}
\end{equation}
whose solution is:

\begin{equation}
\Delta=\frac{1}{2} \frac{\bar \eta}{\left(1-\frac{1}{2N} \right)} \left\{ 1- \frac{\lambda_{\rm GN} }{ \pi } \left(1-\frac{1}{2N} \right)G+\sqrt{\left[1+\frac{\lambda_{\rm GN} }{ \pi }\left(1-\frac{1}{2N} \right)G \right]^2-\frac{2}{N}\left(1-\frac{1}{2N} \right) \left(\frac{\lambda_{\rm GN}}{\pi}\right)^2}\right\}.
\label{PMS3}
\end{equation}
We have taken the positive solution from Eq.~(\ref{PMS2}) since the limit $\lambda \to 0$ of Eq.~(\ref{PMS3}) correctly reproduces the first order in $\delta$ solution displayed in Eq.~(\ref{eta2}). The equation above can be easily expanded in powers of the coupling constant $\lambda_{GN}$:

\begin{equation}
\Delta \simeq \frac{\bar \eta}{\left(1-\frac{1}{2N} \right)} \left\{ 1-  \frac{1}{2N}\left(1-\frac{1}{2N} \right) \left(\frac{\lambda_{\rm GN} }{ \pi } \right)^2 + \frac{1}{2N}\left(1-\frac{1}{2N} \right)^2 \left(\frac{\lambda_{\rm GN}}{\pi}\right)^3 G + {\cal O}\left(\frac{\lambda_{\rm GN}}{\pi}\right)^4 \right\}.
\label{PMS4}
\end{equation}
We can plug $\Delta$ from Eq.~(\ref{Y}) in Eq.~(\ref{PMS4}) and obtain an equation for $\bar \eta$ which, to small values of $\frac{\lambda_{\rm GN}}{\pi}$, reads:

\begin{equation}
{\cal Y}(\eta,\mu_{\uparrow,\downarrow})=\frac{ \pi }{\lambda_{GN}} \frac{1}{\left(1-\frac{1}{2N} \right)}\;,
\label{Y2}
\end{equation}
where ${\cal Y}$ is given by Eq.~(\ref{Y1}). The equation above can not be solved in a closed form for $\bar \eta$ as a function of $\mu_{\uparrow,\downarrow}$. However, for $\delta \mu=0$ i.e., for  $\mu_{\uparrow}=\mu_{\downarrow}=\mu$ it can be easily solved and generalizes Eq.~(\ref{eta0}) for non-zero values of the chemical potential, yielding:

\begin{equation}
\bar \eta_{ \delta^1} (\mu)= \sqrt{\bar \eta_{ \delta^1} (0)(2 \mu - \bar \eta_{ \delta^1} (0))}\;,
\label{eta1}
\end{equation}
where $\bar \eta_{ \delta^1} (0)$ is given by Eq.~(\ref{eta0}), and $\frac{\bar \eta_{ \delta^1} (0)}{2}\leq \mu \leq \bar \eta_{ \delta^1} (0)$. With this equation we find to this order

\begin{equation}
\Delta_{ \delta^1}(\mu)=
\left( 1- \frac{1}{2  N} \right)^{-1} \bar \eta_{ \delta^1} (\mu) \;.
\label{delta1}
\end{equation}

As for $\Delta_{ \delta^1}(0)$ given by Eq.~(\ref{deltaopt}), Eq.~(\ref{delta1}) obtained from the OPT approach, explicitly includes corrections beyond the large $N$ to the mean-field result. Note that the equation above is reduced to the mean-field (local maximum) result, given by Eq.~(\ref{Delta1}), in the limit $N \to \infty$, as it should.

\subsubsection{ The Number Densities at Zero Temperature Beyond Mean-Field}

As we have seen before, the densities are obtained by the usual relation $n_{\uparrow,\downarrow}= -\frac{\partial}{\partial \mu_{\uparrow,\downarrow}} V_{eff}(T,\mu_{\uparrow,\downarrow},\Delta)$ at $\Delta=\Delta_{min} \equiv \overline {\Delta}$. Now since we are considering corrections to the large-$N$ results, we have to include the other minimum condition $\eta = \bar \eta$. Then we want to know how the corrections beyond the mean field will affect the densities given by Eq.~(\ref{n1}). Then one obtains

\begin{eqnarray}
n_{\uparrow,\downarrow} (T,\mu_{\uparrow},\mu_{\downarrow})&=& \frac {1}{v_F \hbar} \left [ (kT)^2 \frac{N}{2 \pi }
I_1^\prime(\eta,\mu_{\uparrow,\downarrow},T)
+ \eta (\eta -\Delta) \frac{N}{2 \pi} I_2^\prime(\eta,\mu_{\uparrow,\downarrow},T)
\right. \nonumber \\
&+& \left.  \frac{\lambda_{\rm GN}}{4 \pi^2} \eta^2 {\cal Y}(\eta,\mu_{\uparrow,\downarrow},T)
I_2^\prime(\eta,\mu_{\uparrow,\downarrow},T)-
\frac{\lambda_{\rm GN}}{4 \pi^2} (kT)^2 I_3(\eta,\mu_{\uparrow,\downarrow},T)
I_3^\prime(\eta,\mu_{\uparrow,\downarrow},T) \right ]
\Bigr|_{\eta = \bar{\eta}, \Delta = {\overline \Delta}}
\label{density}
\end{eqnarray}
where the primes indicate derivatives with respect to $\mu_{\uparrow,\downarrow}$.

Since in this subsection we are interested in obtain the zero temperature corrections to the densities beyond mean-field, the thermal effects will be neglected.

Using Eqs. (\ref{I1T0}), (\ref{I2T0}) and (\ref{I3T0}) in Landau's free energy
density, Eq. (\ref {Vdelta1}), we observe that it can again be divided into two cases: i) $\mu_{\uparrow,\downarrow} < \eta$ and ii) $\mu_{\uparrow,\downarrow} > \eta$. The case $\mu_{\uparrow,\downarrow} < \eta$ is equivalent to the situation where $T=\mu_{\uparrow,\downarrow}=0$ giving $n_{\uparrow,\downarrow} (T=0,\mu_{\uparrow}< \eta,\mu_{\downarrow}< \eta)=0$. For case ii) one has to resort to the numerical solution of the nonlinear equations above.

\section{The Massive Gross-Neveu Model and the Cis-Polyacetylene}

\subsection{The MGN Model Lagrangian}

The cis-polyacetylene has non-degenerate ground states and is appropriately described by the massive Gross-Neveu model (MGN) model. The MGN model is obtained when the discrete chiral symmetry is explicitly broken by adding to the Lagrangian~(\ref{L1}) a fermion bare mass term $M$ for both the spin-$\uparrow$ and spin-$\downarrow$ electrons:

\begin{eqnarray}
{\cal L}_{\rm MGN} =
\sum_{j=1,2} \bar \psi_j [- \gamma_0 \hbar \partial_\tau + i \hbar v_F \gamma_1 \partial_x -  \Delta(x) - M +  \gamma_0 \mu_j  ] \psi_j -\frac{1}{ \hbar v_F \lambda_{\rm GN}}  \Delta^2(x).
\label{LagMGN}
\end{eqnarray}
The bare mass term $M$ corresponds~\cite{gnpolymers2} to the Brazovskii and Kirova's external contribution from the rigid polymer skeleton to the total gap~\cite{CB1}. As we will see below, the presence of a (bare) mass term ensures the elimination of kink-antikink configurations, which are suppressed in the thermodynamical limit even in presence of finite chemical potentials for the electrons~\cite{Barducci}. Then in this section we examine only homogeneous condensates. In presence of a mass term it is more convenient to redefine the $\Delta$ field by shifting it by a constant $\Delta \to \Delta + M$. Then the Lagrangian (apart from constant terms) reads:

\begin{eqnarray}
{\cal L}_{\rm MGN} =
\sum_{j=1,2} \bar \psi_j [- \gamma_0 \hbar \partial_\tau + i \hbar v_F \gamma_1 \partial_x -  \Delta +  \gamma_0 \mu_j  ] \psi_j -\frac{2M}{ \hbar v_F \lambda_{\rm GN}}  \Delta -\frac{1}{ \hbar v_F \lambda_{\rm GN}}  \Delta^2.
\label{LagMGN2}
\end{eqnarray}

\subsubsection{The Effective Potential of the MGN Model at Zero Temperature and Zero Chemical Potentials}

Proceeding as before, we obtain the following expression for the renormalized effective potential of the MGN model at zero temperature and chemical potentials

\begin{eqnarray}
\label{poteffMGN}
V_{eff}(\Delta, M)=\frac{\Delta^2}{ \hbar v_F } \left(\frac{1}{\lambda_{\rm GN}} - \frac{3}{2 \pi} \right) + \frac{\Delta^2}{\pi \hbar v_F } \ln \left( \frac{\Delta}{m_F} \right) + \frac{2M}{ \hbar v_F \lambda_{\rm GN}}  \Delta.
\end{eqnarray}
Substituting $m_F$ from the solution for $\Delta$ when $M=0$ we find: 

\begin{eqnarray}
\label{poteffMGN2}
V_{eff}(\Delta,M)=\frac{\Delta^2}{2 \pi \hbar v_F } \left[ \ln \left( \frac{\Delta^2}{\Delta_0^2} \right)-1 \right] + \frac{2M}{ \hbar v_F \lambda_{\rm GN}}  \Delta.
\end{eqnarray}

The minimization for $V_{eff}(\Delta,M)$ gives:

\begin{equation}
\Delta \ln\left(\frac{\Delta^2}{\Delta_0^2} \right) + \frac{2\pi M}{\lambda_{\rm GN}}=0.
\label{MinV1}
\end{equation}
From the equation above it is easy to see that when $M=0$ the solutions are $\Delta=0$ and $\Delta = \pm \Delta_0$ at $x = \pm \infty$, interpolated by a kink, as seen in Section~\ref{SB}. Besides, we have that $V_{eff}(\Delta=0)=0$ and $V_{eff}(\Delta=\pm \Delta_0)=-\frac{\Delta_0^2}{2 \pi \hbar v_F }$. Then, since $\Delta=0$ corresponds to a local maximum, the minima is degenerate at $\pm m_F e^{ 1- \frac {\pi}{\lambda_{\rm GN} }}$. For any $M \neq 0$ the degenerate vacuum (minima) disappear, as well as the kink, giving place to a unique solution for a given value of $M$~\cite{Feinberg2}. As pointed out in Ref.~\cite{Feinberg2} if any soliton exists at all, its stability has to depend on the energetics of trapping fermions. 

Eq.~(\ref{MinV1}) is very sensitive to the value of $M$. In other words, there will be solutions for (positive) $\Delta$ only up to a maximum value of $M$. If we define a dimensionless function $f(x, \bar M)\equiv f(x)= x^2 \ln\left(x^2 \right) + \bar M x$, where $x \equiv \frac{\Delta}{\Delta_0}$ and $\bar M \equiv  \frac{2 \pi M}{\lambda_{\rm GN} \Delta_0}$, we have to solve self-consistently $f(x)=0$ and $f'(x)=\frac{df(x)}{dx}=0$ to obtain the windows for $M$ and $\Delta_0(M)$:

\begin{equation}
0 \leq M \leq \frac{\lambda_{\rm GN} \Delta_0}{ \pi e},~~~~~~\frac{\Delta_0}{e} \leq \Delta_0(M) \leq \Delta_0.
\label{MinV2}
\end{equation}
The effect of a mass term in the MGN model can be summarized simply as: as $M$ is increased from $0$ to $M_{max}=\frac{\lambda_{\rm GN} \Delta_0}{ \pi e}$, the minimum of $V_{eff}(\Delta,M)$ moves smoothly from $\Delta_0$ to $\frac{\Delta_0}{e}$.

\subsubsection{The Effective Potential of the MGN Model at Zero Temperature and Finite Chemical Potentials}

Since after the change of variable $\Delta \to \Delta + M$ the mass parameter $M$ enters only in the classical part of the effective potential, it can be written as

\begin{eqnarray}
V_{eff}(\Delta, M, \mu_{\uparrow,\downarrow})=V_{eff}(\Delta, M) + \Theta_1 {\cal F}_{1} + \Theta_2 {\cal F}_{2},
\label{poteffMGN3}
\end{eqnarray}
where $V_{eff}(\Delta, M)$ is given by Eq.~(\ref{poteffMGN2}) and ${\cal F}_{1,2}$ are defined after Eq.~(\ref{potefff}). The extremization of $V_{eff}(\Delta, M, \mu_{\uparrow,\downarrow})$ yields

\begin{equation}
\label{minM}
\Delta \left[\ln \left( \frac{\Delta^2}{\Delta_0^2} \right) +  \Theta_1 {\cal G}_1 + \Theta_2 {\cal G}_2 \right]+ \frac{2\pi M}{\lambda_{\rm GN}}=0,
\end{equation}
where ${\cal G}_{1,2}\equiv \ln \left(\frac{\mu_{\uparrow,\downarrow} + \sqrt{\mu_{\uparrow,\downarrow}^2-\Delta^2}}{\Delta} \right)$. In the limit $M \to 0$ Eq.~(\ref{minM}) reduces to Eq.~(\ref{min2}). The equation above can be solved only numerically for arbitrary values of $\mu_{\uparrow,\downarrow}$ and $M$. However, for $\delta \mu =0$ we obtain the following $\mu$ and $M$ dependent equation to first order in $M/\Delta$:

\begin{equation}
\label{minM2}
\Delta^3 + (\Delta_0^2 -2 \Delta_0 \mu)\Delta -  \frac{2\pi M}{\lambda_{\rm GN}}(\Delta_0 -\mu)\Delta_0=0.
\end{equation}
This equation has the analytical solutions:
\newline
{\bf 1.} When $\mu=\Delta_0/2$, $\Delta=\Delta_0 \left(\frac{\pi M}{\lambda_{\rm GN} \Delta_0} \right)^{1/3}$;
\newline
{\bf 2.} When $\mu=\Delta_0$, $\Delta=0$. The other solution, $\Delta =\Delta_0$, is a local maximum. This can be understood simply as: for small $M$ and such a large value of $\mu$ the broken symmetry has been restored.
\newline
For other values of $\mu$, Eq.~(\ref{minM2}) has to be solved numerically.

\section{Summary of the Results}

We show in Tab.~\ref{tableone} several physical quantities calculated within the GN model at mean-field and OPT (beyond mean-field) approximations at various medium conditions, i. e., temperature and densities. The places in the table with the $?$ means that, to the best of our knowledge, there are no studies concerning that physical quantities at mean-field and beyond in that situations.

\section{Conclusions}
We have studied the influence of an external Zeeman magnetic field $B_0$ on 1D $\rm (CH)_x$ chains, which can be converted into a partially or fully polarized organic conductor, depending on the intensity of the $B_0$ applied on the wire. The mean-field finite temperature phase diagram of the chain has been investigated. We found that the gap parameter and the magnetization (as well as the densities and density imbalance, and the magnetic susceptibility) of this 1D material under $B_0$ have a similar second-order behavior until their respective tricritical points are reached. Below these points the transitions are of first order. We found these two tricritical points analytically.

He have shown that the location of the tricritical point in the $t$ $\left(=\frac{T}{T_c(0)}\right)$ versus $\nu$ $\left(=\frac{7 \xi(3)}{8 \pi^2} \frac{(\mu_{\uparrow}^2+\mu_{\downarrow}^2)}{(k_B T_c(0))^2}\right)$ phase diagram stays at the same place by considering both homogeneous and inhomogeneous condensates, as occur in symmetric $\delta \mu = B_0 = 0$ systems~\cite{gnpolymers,gnpolymers2,gnpolymers3,Basar}. We have found the critical magnetic field $B_{0,c}$ necessary to have a fully polarized organic conductor at zero and finite temperature. We found that for a given magnetic field below $B_{0,c}$, partial polarizations (magnetizations) can be realized experimentally, provided the temperatures are kept outside the ``$non-metallic$" region of Fig.~\ref{pd1}. It is worth noting that, according to Eq.~(\ref{cmf}), for other 1D systems with a smaller critical chemical potential or with a greater effective $g$-factor, a smaller critical magnetic field necessary for a fully polarization of these 1D chains would be found.

In Table \ref{tableone} we show the homogeneous and inhomogeneous gap parameter $\Delta(x)$, the Pauli magnetization $M$ and the magnetic susceptibility $\chi$, and zero and finite $T$ and $\mu$ calculated at mean-field approximation and beyond. The quantities identified with the $?$ mean that further investigation are needed to find them.

We have also seen that the consideration of electron correlations in the investigation of some specific properties of 1D ${\rm CH}$ chains is of crucial importance. Then we intent to analyze a continuum version of the Hamiltonian of Eq.~(\ref{PPP2}) within the quantum field theory approach, and publish it elsewhere.

A direct extension of the present work would be the investigation of the possibility of using polymers or graphene ribons as spin-polarized conductors, upon the joint application of a Zeeman and an electric field in the system. This would enable to study the transport properties of 1D chains at zero and finite temperature.

As a next step in the investigation of the effects of magnetic fields on 1D systems, we would like to study the results of parallel and perpendicular magnetic fields applied simultaneously and independently on a polymer or graphene wire. This could open interesting possibilities for using these type of materials in practical applications as electronic devices.

The methods we used here, based on the calculation of the (grand) effective potential free energy, which are more suitable for the analysis of the phase structure of the model, are useful also for the study of the consequences of a perpendicular magnetic field applied in the system. Indeed, one can incorporate the effects of a magnetic background on the effective potential and investigate how the chiral transition and the condensate (in our case the $\Delta$ gap) are modified. Assuming that the system is now in the presence of a strong magnetic field background that is constant and homogeneous, one can compute the modified effective potential following the procedure presented in detail in Refs. \cite{Fraga1,Fraga2,Marcus,Fraga3}. We hope to make further studies on the magnetic properties of these systems based on the model we have studied here, and to report on them in the future.

\begin{table}[t]
\begin{center}
\begin{tabular}{|l| c c c c c|}
\hline
&&&&&\\[-2mm]
Approximation $\setminus$ Physical quantity
& Homogeneous $\Delta$ & Inhomogeneous $\Delta(x)$ &  M & $\chi$
\\
&&&&&\\[-2mm]
\hline
&&&&&\\[-2mm]
MF at zero $\mu$ and zero $T$ & $\Delta_0=m_F e^{ 1- \frac {\pi}{\lambda_{\rm GN} }}$ & $\Delta_0 \tanh(\Delta_0 x)$ & $0$ & $0$  \\
MF at finite $\mu$ and zero $T$  & $\Theta(\mu_c-\mu) \Delta_0$ & * & $0$ & $0$ \\
&&&&&\\[-2mm]
MF at finite $\mu_{\uparrow,\downarrow}$ and zero $T$ & Num. Solut. of Eq.~(\ref{min2}) & ? & $ \frac{2 \mu_B^2}{ \pi \hbar v_F} B_0$ & $ \frac{2 \mu_B^2}{ \pi \hbar v_F} $ \\
&&&&&\\[-2mm]
MF at finite $\mu_{\uparrow,\downarrow}$ and high $T$ & $ \sqrt{-\frac{\alpha_2}{2 \alpha_4}}$ & $ \Delta_0(\varsigma) \sqrt{\varsigma} ~{\rm sn}(\Delta_0 x; \varsigma )$ & Eq.~(\ref{mag2}) &  Eq.~(\ref{ms1})  \\
&&&&&\\[-2mm]
Beyond MF at zero $T$ and zero $\mu$   & $F(N)^{-1} m_F  e^{-\frac {\pi}{\lambda_{\rm GN}F(N) }}$ & ? & ? & ?  \\
&&&&&\\[-2mm]
Beyond MF at finite $\mu_{\uparrow,\downarrow}$ and  zero $T$  & Num. Solut. of Eq.~(\ref{Y2}) & ? & ? & ? \\
&&&&&\\[-2mm]
Beyond MF at finite $\mu_{\uparrow,\downarrow}$ and $T$   & ? & ? & ? & ? \\
\hline
\end{tabular}
\end{center}
\caption{\small Physical quantities calculated at MF and beyond approximations and situations for the temperature and chemical potentials. The inhomogeneous gap parameter at zero $T$ and finite $\mu$ marked with the $*$ symbol is discussed below Eq.~\ref{difeq3}. $\Delta_0(\varsigma)$ is given by Eq.~(\ref{difeq4}), $F(N)=1-1/2N$, $M$ is the Pauli magnetization and $\chi$ is the magnetic susceptibility which are, of course, zero for $\delta \mu =\frac{g}{2}\mu_B B_0=0$.}
\label{tableone}
\end{table}  

\begin{center}
Acknowlegments
\end{center}
This work has benefited greatly from many conversations with M. Thies. I also would like to thank A. L. Mota and R. O. Ramos for enlightening discussions. The author acknowledges partial support by the Brazilian funding agencies Conselho Nacional de Desenvolvimento Cient\'{i}fico e Tecnol\'{o}gico (CNPq) and Funda\c{c}\~{a}o de Amparo a Pesquisa do Estado de Minas Gerais (FAPEMIG).

\end{document}